\documentclass[12pt,letter]{article}
\usepackage{graphicx}
\usepackage{amsmath}
\usepackage{amssymb}
\usepackage{authblk}
\usepackage[left=1in,right=1in,top=1in,bottom=1in]{geometry}
\usepackage{setspace}
\usepackage{float}
\usepackage[super,sort&compress,comma]{natbib}
\usepackage[font={footnotesize}]{caption} 
\usepackage[labelfont=bf]{caption}
\usepackage{bm}
\usepackage[T1]{fontenc}
\usepackage{microtype}
\usepackage{subfigure}
\usepackage{chemformula}
\usepackage[version=4]{mhchem}
\usepackage{textcomp}
\usepackage{color}
\usepackage{mathtools}
\usepackage{gensymb}
\usepackage{esvect}
\usepackage{booktabs}
\usepackage{multirow}
\usepackage{indentfirst}
\usepackage{mathrsfs}
\usepackage{float}
\usepackage{hyperref}
\usepackage{xcolor}
\usepackage{booktabs}
\usepackage{amssymb} 
\usepackage{bm} 

\begin{document}

\title
{\textbf{A Workflow to Create a High-Quality Protein-Ligand Binding Dataset for Training, Validation, and Prediction Tasks}}

\author{Yingze Wang$^{1,\Delta}$, Kunyang Sun$^{1,\Delta}$, Jie Li$^{1}$,
Xingyi Guan$^{1}$, Oufan Zhang$^{1}$, Dorian Bagni$^{1}$, 
Yang Zhang$^{4-6}$, Heather A. Carlson$^{7}$, Teresa Head-Gordon*$^{1-3}$}
 \date{}
\maketitle
\begin{center}
\vspace{-10mm}
$^1$Kenneth S. Pitzer Theory Center and Department of Chemistry, $^2$Department of Bioengineering, $^3$Department of Chemical and Biomolecular Engineering, University of California, Berkeley, CA, 94720 USA

$^4$Department of Computer Science, School of Computing, National University of Singapore, 117417,  $^5$Cancer Science Institute of Singapore, National University of Singapore, 117599, $^6$Department of Biochemistry, Yong Loo Lin School of Medicine, National University of Singapore, 117596, Singapore

$^7$Odyssey Therapeutics Inc. 1350 Highland Dr., Ann Arbor, MI, 48108, USA

$^{\Delta}$authors contributed equally

corresponding author: thg@berkeley.edu
\end{center}

\begin{abstract}
\noindent
Development of scoring functions (SFs) used to predict protein-ligand binding energies requires high-quality 3D structures and binding assay data for training and testing their parameters. In this work, we show that one of the widely-used datasets, PDBbind, suffers from several common structural artifacts of both proteins and ligands, which may compromise the accuracy, reliability, and generalizability of the resulting SFs. Therefore, we have developed a series of algorithms organized in a semi-automated workflow, HiQBind-WF, that curates non-covalent protein-ligand datasets to fix these problems. We also used this workflow to create an independent data set, HiQBind, by matching binding free energies from various sources including BioLiP, Binding MOAD and BindingDB with co-crystalized ligand-protein complexes from the PDB. The resulting HiQBind workflow and dataset are designed to ensure reproducibility and to minimize human intervention, while also being open-source to foster transparency in the improvements made to this important resource for the biology and drug discovery communities.
\end{abstract}


\section{INTRODUCTION}
\noindent
Scoring functions (SFs) are crucial in computer aided drug discovery, utilized for selecting the most probable ligand geometry and its binding pose with a protein that best correlates or predicts their free energy of binding.\cite{SF_evaluation} There are a plethora of SFs being developed and widely used by computational and medicinal chemists, and they can be broadly categorized into either classical scoring functions\cite{autodock_vina, autodock4, glide1, glide2, OpenFF1, OpenFF2, xscore, GOLDscore, PMF, mmscore, drugscore} or machine learning scoring functions.\cite{rfscore,IGN,pignet,rtmscore,deepDTA,holoprot,tankbind,GIGN} The majority of protein-ligand SF predictors, whether physical or machine-learned, have been trained on the PDBbind dataset\cite{pdbbind_2004, pdbbind_2005, pdbbind_2009, pdbbind_2014_1, pdbbind_2014_2, pdbbind2014, pdbbind_2016} (http://www.pdbbind-cn.org/), specifically v2020, a curated set of $\sim$19,500 biomolecular complex structures and their experimentally measured binding affinities. PDBbind is further organized into a "general" data subset that is often adopted by SFs for training, and separate "refined" and "core" datasets which contain protein-ligand complexes with the best structural quality and most reliable binding affinity data that is used for testing. Various benchmarks based on PDBbind, such as CASF (Comparative Assessment of Scoring Functions) series\cite{casf2016,casf2009,casf2013,casf2013_2}, CSAR (Community Structure Activity Resource) 2010\cite{csar2010} and PDBbind-blind-2013\cite{pdbbindblind} have been proposed to assess the scoring power, ranking power, docking power and screening power of various SFs.

PDBbind has been an invaluable resource to the biomolecular community during its two-decade development, but a significant portion of the PDBbind dataset contains structural errors, statistical anomalies, and a sub-optimal organization of protein-ligand classes that can limit SF training and validation\cite{lppdbbind,equibind}. These inconsistencies undermines the purpose of the refined set, which is intended to serve as a high-quality benchmark for evaluation of scoring functions and docking methods. Another concern in regards PDBbind is that the data processing procedure is neither open-sourced nor automated, potentially relying on individual groups needing to introduce their own manual intervention that may lead to inconsistencies. Furthermore, the PDBbind data curation process became more problematic in 2021 when PDBbind ceased to be freely available for data curated after 2020, which limits access and hinders the development and validation of new scoring functions (and other additional uses). 

Fortunately, other curation efforts have created alternative protein-ligand structural and/or binding datasets that have increased the size and comprehensiveness of available data for drug discovery efforts. BindingDB is a database containing 2.9 million binding measurements spanning 1.3 million compounds for thousands of protein targets, which are curated from the literature and patents.\cite{bindingdb1,bindingdb2,liu2025bindingdb} Binding MOAD is a curated database of 41,409 protein–ligand structural complexes, with binding affinity data available for 15,223 (37\%) of them; Binding MOAD's curation involved extracting high-quality structures from the PDB and finding associated binding data from publications with the aid of an NLP-based annotation tool.\cite{moad2005,moad2008,moad2023,BindingMOAD} BioLiP is a large database of over 900,000 biologically-relevant protein-ligand interactions curated from the PDB, and enriched with various functional annotations, including Enzyme Commission numbers, Gene Ontology terms, catalytic sites, and binding affinities from Binding MOAD, BindingDB, as well as manual surveys.\cite{biolip,biolip2} Other related datasets that focuses more on the geometries of proteins and ligands, including PLINDER\cite{durairaj2024plinder} and DockGen\cite{dockgen}, contain an expanded set of protein-ligand structural complexes but do not have annotations of binding affinity data. However, in general, these curation efforts have largely focused on increasing the size and comprehensiveness of protein-ligand data, rather than increasing the quality and reliability of the data themselves. Therefore, there is a pressing need for an open-source and systematic workflow to prepare protein-ligand binding datasets with well-defined binding affinity annotations and higher-quality structures in order to foster greater reproducibility, transparency, and accessibility.

In this work, we introduce HiQBind-WF, a workflow of algorithms for data cleaning and structural preparation that creates a curated dataset of high-quality, non-covalent protein-ligand complex structures with binding affinity annotations. This workflow contains several modules: (1) a curating procedure that rejects ligands covalently bonded to proteins, ligands with rarely-occurring elements, and structures containing severe steric clashes; (2) a ligand-fixing module to ensure the correctness of the ligand structure including correct bond order and reasonable protonation states; (3) a protein-fixing module to extract and, when necessary, add missing atoms to all chains involved in the protein-ligand binding; (4) a structure refinement module to simultaneously add hydrogens to both proteins and ligands in their complex state, as opposed to the current practice in PDBbind that completes the hydrogen chemistry for protein and ligand independently. The motivation for adding this hydrogen growth module is that although many SFs only take heavy atoms into consideration, future physics-based SFs could potentially benefit from explicit hydrogens to better model intermolecular interactions such as hydrogen bonding. 

We utilized this workflow to optimize PDBbind v2020 and compared the processed structures. Analysis of the structural differences between the same PDB entry demonstrated that HiQBind-WF is able to correct for various observed structural imperfections. Further, to illustrate the applicability of the HiQBind-WF, we created HiQBind, a new dataset with high-quality protein-ligand binding structures and affinities by processing PDB entries included in BioLiP2 and Binding MOAD associated with binding affinities drawn from BindingDB. The HiQBind dataset includes >18,000 unique PDB entries and >30,000 protein-ligand complex structures. We also confirmed that HiQBind shares similar properties with existing datasets like PDBbind, demonstrating its feasibility to be used for developing and validating SFs and other structure-based drug-design tools. The HiQBind-WF and HiQBind dataset are provided open-source to foster transparency and sustainability as new data appears, in order to maintain this important resource for the biology and drug discovery communities.

\vspace{-3mm}

\section{METHODS}
\begin{figure}[H]
\begin{center}
\includegraphics[width=0.9\textwidth]{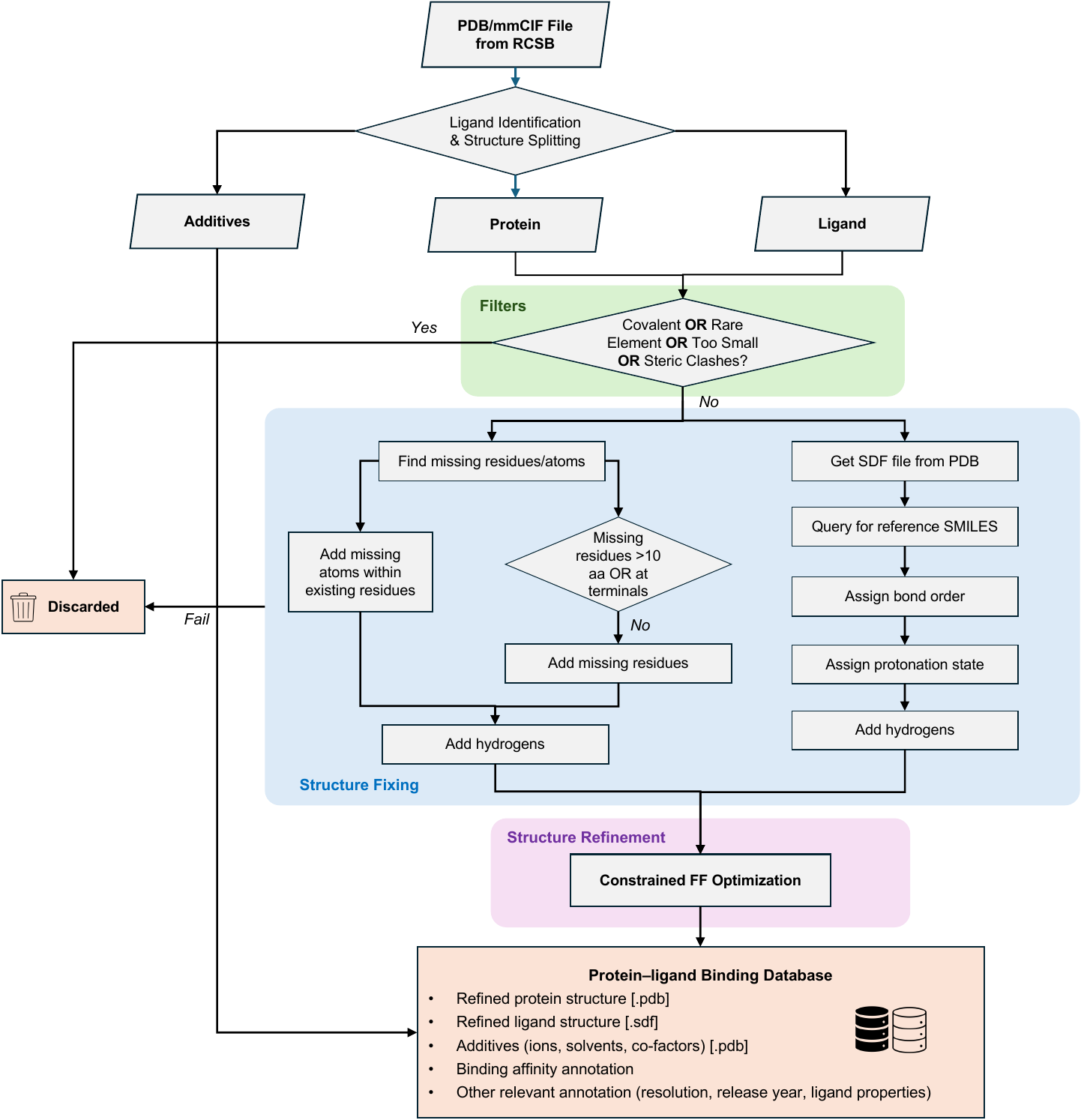}
\end{center}
\caption{\textit{Schematic representation of the semi-automated HiQBind-WF to refine protein-ligand binding structures.} HiQBind-WF downloads the pdb and mmCIF files from the RCSB PDB\cite{Berman2000}, followed by splitting each structure into three components —ligand, protein, and additives. A series of filters are then applied to remove covalent binders, ligands with rare elements, very small ligands, and complexes exhibiting steric clashes. Subsequently, the protein structure is fixed by adding missing atoms and residues (ProteinFixer) and the ligand structure is fixed by correcting bond orders, protonation states, and aromaticity (LigandFixer). Finally, the fixed protein and ligand structures are recombined and subjected to a constrained energy minimization to resolve potential unreasonable structures and to refine hydrogen positions.}
\label{fig:wf}
\end{figure}

\noindent
The flowchart of HiQBind-WF is illustrated in Figure \ref{fig:wf}. We start by downloading the pdb and mmcif formats directly from the RCSB PDB\cite{Berman2000} for supplied entries. The pdb files are used for structure preparation and the headers in mmcif files are used to extract useful metadata, such as resolution, deposit date and sequence information. For each PDB entry, we split the structure into three components: ligand, protein and additives, and curate these categories as follows. 

We define three classes of ligand(s) for any given protein-ligand complex structure: (1) Any residue will be identified as a ligand if its name matches the Chemical Component Dictionary (CCD) code deposited in given reference datasets (PDBbind, BioLiP or Binding MOAD). Ligands identified in this manner are referred to as "small molecules". (2) Otherwise, chains in the original PDB file that are less than 20 residues but more than one residue as ligands will be selected as ligands. For example, PDBbind entries that contain patterns such as “*-mer,” or symbols like "-", "\&" or "+", MOAD entries with more than one CCD in the \texttt{name} column of its csv-formatted dataset, and BioLiP entries with \texttt{Ligand CCD} column to be "peptide", "dna" or "rna".  These ligands are typically polypeptides, oligosaccharides, or oligonucleotides, collectively referred to as "polymers". (3) For each identified ligand, we label any biopolymer chains within $\mathrm{10\ \mathring{A}}$ as the associated protein structure. Then, for each protein structure, we labeled residues specified by the "HETATM" record in the pdb file within  $\mathrm{4\ \mathring{A}}$ as additives, which includes ions, solvents, and co-factors. The additives are saved in pdb format and directly deposited in the database, and the protein and ligand structure are ready to proceed to the next workflow steps.

After the splitting for the ligand categories and their  associated protein-ligand complexes, we define an additional set of downselect filters, including some borrowed from the processing protocols of LP-PDBbind\cite{lppdbbind}. The purpose of these filters are to exclude protein-ligand complex structures that specifically can interfere with training of SFs, with eliminations that would meet any of the following criteria:

\begin{itemize}
    \item \underline{Covalent binder filter}: Excludes ligands covalently bound to the protein, as indicated by the “CONECT” record in the pdb file.  When covalently-bound ligands are identified, they are eliminated. This is because covalent binding inherently is different from non-covalent binding which does not involve bond breaking, and thus requires special treatment in any SFs. Those covalent binder entries are included in the SI which may be helpful as a separate curation of the data or may be accessible from CovBinderInPDB\cite{guo2022covbinderinpdb}.
    \item \underline{Rare element filter}: Excludes ligands containing elements other than H, C, N, O, F, P, S, Cl, Br, I. For example, Te or Se are infrequently encountered, and their inclusion can make it challenging for SFs to learn key binding features giving data sparsity for these ligands. These ligand entries are also included in the SI which may be helpful as a separate curation of the data.
    \item \underline{Small ligand filter}: Excludes ligands containing less than 4 heavy atoms, which includes small inorganic binders like $\mathrm{O_2,NH_3,CO_2,NO_2,N_3^{-}}$, which are  beyond the scope of common protein-ligand binding studies.
    \item \underline{Steric clashes filter}: Excludes structures with protein-ligand heavy atom pairs closer than 2 angstroms. Such steric clashes often arise from electron density uncertainties or inaccurate structural reconstruction from electron densities and are not physically feasible non-covalent interactions. Including such structures in SF development could be detrimental, for example leading to an underestimation of the repulsion energy in physics-based SFs. Additionally, the steric clash filter helps to exclude covalent ligands if the covalent bond is not properly represented in the "CONECT" record.
\end{itemize}

For protein-ligand complexes that pass these filters, two structure-fixing modules are implemented separately for proteins and ligands. In the ProteinFixer module, we first use the sequence information from the mmcif file header to detect missing atoms and residues. Then, for missing residues or missing atoms within an existing residue, PDBFixer\cite{pdbfixer} (version 1.9) is used to add them, except when the missing residues are longer than 10 amino acids or are located at the sequence terminals. Adding missing atoms to protein structure is essential near binding sites because incomplete structures can compromise accurate modeling of binding interactions, and any molecular dynamics or alchemical binding free energy calculation also require complete structures to ensure the correct structural ensemble are sampled during simulations. However, long missing segments or missing terminus residues in crystal structure are often attributable to intrinsically disordered regions (IDR)\cite{liu2024}, domains that are not expressed in the samples for crystallization, or his-tags introduced in the protein purification process\cite{idr_in_pdb_2007}. If far enough removed from the ligand binding site(s), we regard it safe to skip modeling these residues explicitly. Hence, we leave these regions in their original form and in themselves do not define a criterion for being discarded in the final dataset. The final step of the protein-fixing module is to add hydrogen atoms at pH=7.4 with PDBFixer. At this pH, the protonation state assignment of titratable side-chains obeys the following rules: all lysine (LYS) and arginine (ARG) are positively charged and glutamic acid (GLU) and aspartic acid (ASP) are deprotonated. Histidine remains in a neutral form and whether the HID or HIE variant (the hydrogen is added to N$\delta$ or $N\epsilon$, respectively) is selected will be based on which one forms a better hydrogen bond, which is the default behavior of PDBFixer.
 
In the LigandFixer module, we first obtain an sdf file for each ligand instance either by downloading from the RCSB PDB (if possible) or converting from the native pdb format with OpenBabel\cite{openbabel}. Since explicit atom connections may not be present in the pdb format, the bond orders in this converted sdf file are typically inferred from local atomic geometries and the resulting structure is herein referred to as "inferred structure". Then, a reference SMILES is obtained, which is used to correct bond orders and aromaticity specifications that could sometimes be mislabeled in the inferred structure. The bond order assignment protocol is implemented as follows: if 
the inferred and reference structure are isomorphic, a one-to-one atom mapping will be generated by structure matching and then bond orders, atom hybridization and aromatic specifications will be assigned according to the reference. Otherwise, the bond order assignment will come to a failure point, which means that the inferred structure does not share the same number of atoms or bond connectivity as the reference, indicating that there are missing atoms or distorted geometries in the crystal structure. Therefore, such structures will be excluded. 

After a correct structure is obtained, protonation states are assigned to the ligand. We acknowledge that it is a non-trivial task to correctly determine protonation states for titratable groups within a ligand at a given pH and many algorithms that use empirical rules, QM/MM calculations or machine learning have been reported\cite{unipka,epik7,chemaxon}. However, since our workflow is designed for high-throughput processing, we improve the efficiency using a simple set of predefined rules to determine the protonation states by relevant matching functional groups in SMARTS patterns. Acids, nitro groups, thiophenols, azides, and N-oxides are deprotonated. Aliphatic amines and guanidines/imines are protonated, while anilines are not protonated. There are other special considerations that should also be accounted for: amines will not be protonated if the nitrogen is directly bonded with atoms other than H or C. Only one nitrogen atom on diamines and piperaizines will be protonated to avoid two positive charged groups close to each other, which is not favorable at normal biologically-relevant pH. Enols with the motif O=C-C=C-OH are deprotonated. The protonation state assignment is implemented by modifying the default behavior of the \texttt{dimophite\_dl} package\cite{dl} which can be found in the Github repository.

One thing that should be noted here is the source of the reference SMILES string. If the ligand is a small molecule with a CCD code or is a polymer with a BIRD (Biologically Interesting Molecule Reference Dictionary\cite{Berman2000}) code, we will query RCSB PDB for its reference SMILES. If the ligand is a polymer consisting only of alpha-amino acids, we will assume it is a simple non-cyclic peptide and generate a SMILES string based on its sequence information and amide-bond formation rules. Apparently, for the latter case, any mismatch between the inferred and reference structure does not mean the inferred structure is wrong - the ligand may just be a cyclic peptide or contain disulfide bonds. However, such structures will also be excluded as we are unable to verify its correctness automatically at this stage. For such cases, human inspection will be inevitable and it's beyond the scope of the workflow. In addition, we found that some of the SMILES strings deposited in RCSB PDB are incorrect such that all the bonds are labeled as single bonds. Most of these errors were caught by a geometric check for sp$^3$/sp$^2$/sp carbons. For these cases, we manually corrected the SMILES according to the original literature and use the corrected one to do the ligand fixing. The list of manually corrected SMILES can be found in the public Github repository. The bond-order assignment, protonation state assignment, and added hydrogens in the ligand-fixing module are all performed with RDKit\cite{rdkit} (version 2024.03.4).

The last part of the HiQBind-WF structure preparation is a refinement module in which the fixed ligand structure and protein structure are combined, followed by a constrained energy minimization with a well-established force field. AMBER14SB\cite{amber14} is used for the protein and OpenFF-2.1.0\cite{OpenFF2} together with Gasteiger charges\cite{Gasteiger1978ANM} are used for the ligand. Coordinate constraints are applied to all atoms that are experimentally resolved, which means only positions of hydrogens (both on the ligand and protein) and atoms added by PDBFixer in the protein-fixing module are allowed to be optimized. We found this physically-based structural optimization is useful to resolve any remaining steric clashes between added atoms introduced by treating the protein and ligand structure separately in the previous structure fixing modules. Additionally, the hydrogen-bonding network between the ligand and protein is also optimized in this process. The constrained energy minimization was performed with OpenMM 8.1.1\cite{openmm8} by setting masses of all constrained atoms to zero. 

The binding affinity in terms of $\Delta G$ is directly related to the dissociation coefficient $K_d$ or $K_i$ through the standard relationship $\Delta G=RT \ln(K_{d/i})$.\cite{thermodynamics} However, a large portion of the data in the binding datasets is reported in terms of $\mathrm{IC}_{50}$, which cannot be easily translated to $\Delta G$s due to its dependence on other experimental conditions and inhibition mechanisms.\cite{activity_relationship} Furthermore, the $\mathrm{IC}_{50}$ values for the same protein-ligand complex can vary up to over order of magnitude in different assays\cite{ross2023maximal}, and some deposited binding data are not reported as exact values but just ranges. Therefore, the binding affinity data is reorganized into a machine-readable format (csv) with comments as to the form of the experimental binding free energy data: $K_d$, $K_i$, $\mathrm{IC}_{50}$ and $\mathrm{EC_{50}}$.

\section{RESULTS}
\begin{figure}[H]
\begin{center}
\includegraphics[width=1.0\textwidth]{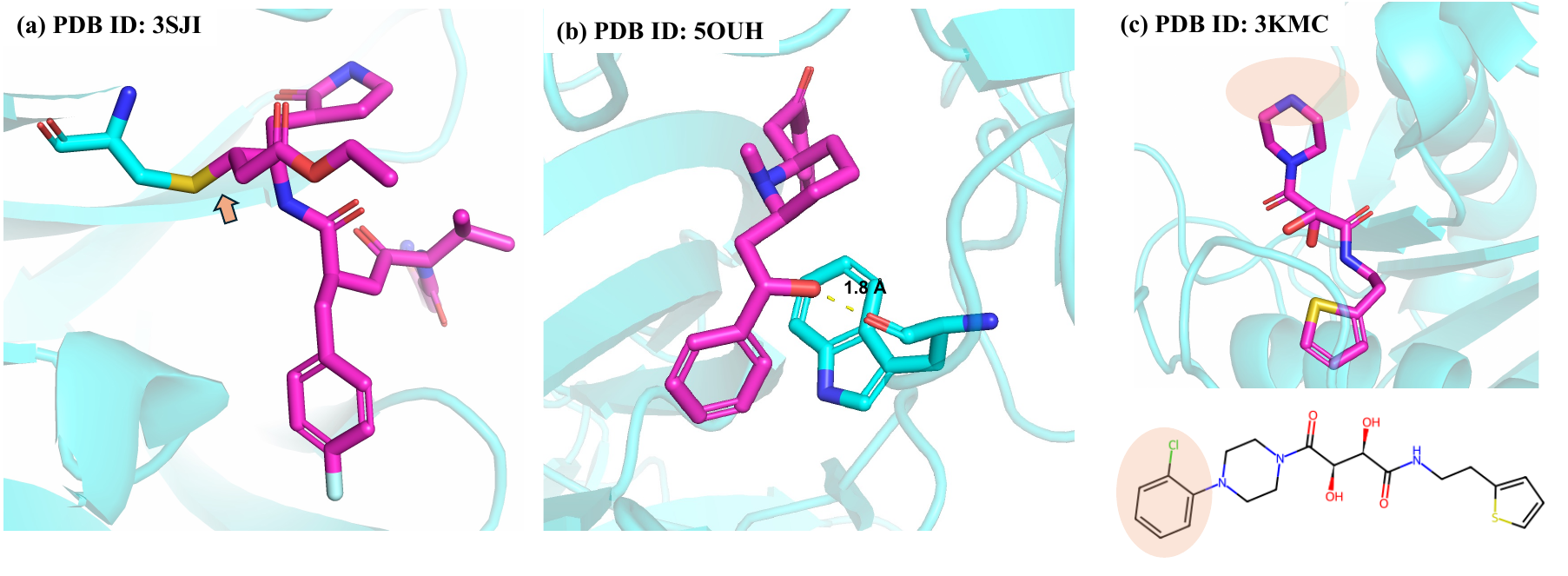}
\end{center}
\caption{\textit{Common structural imperfections in PDBbind dataset.} (a) Covalent binders. The ligand is covalently bound to cysteine with a Michael addition reaction. (b) Steric clashes with the distance between the clashing atoms being only 1.8 Å. (c) Missing atoms.}
\label{fig:err}
\end{figure}

\noindent
Figure \ref{fig:err} illustrates the common problems that arise in the training and tests sets for protein-ligand interactions and associated binding assays when developing a scoring function using various curated databases. Some of the structural imperfections are inherited from the original RCSB Protein Data Bank (PDB)\cite{Berman2000} dataset, such as missing hydrogen atoms and/or incomplete residues due to uncertainties in the modeled electron densities, whereas some errors originate from the preparation of ligand structures that results in incorrect bond order, protonation state, tautomer state and aromaticity specifications. Some entries are covalent binders such as shown in Figure \ref{fig:err}a, which requires special methods to account for the covalent bond formulation\cite{covdock,autodock_cov}, and should remain distinct from protein-ligand complexes that are formed from non-covalent interactions only. Figure \ref{fig:err}b illustrates an example from the PDBbind refined set,   5OUH\cite{5ouh}, which is a noncovalent binder that exhibits a severe atomic clash with the protein. Similarly, for 3KMC\cite{3kmc} the chlorobenzene portion of the ligand is absent from the crystal structure as seen in Figure \ref{fig:err}c. 
This again highlights the need for the community to have a free open-source tool to curate high-quality protein-ligand structures in a reproducible way.

\noindent
\subsection{The HiQBind Workflow}  
\noindent
We applied HiQBind-WF to refine the structures in the publicly available PDBbind v2020 dataset\cite{pdbbind_2016}. While we are not able to publish this optimized PDBbind dataset because the user's agreement of PDBbind prohibits any distribution of any derivative dataset, we can report some general statistics of the workflow and provide examples of structural fixes of the protein and ligand data. However, users can reproduce an optimized PDBbind data set using HiQBind-WF following step-by-step instructions in our Github repository.

Of the original 19,443 unique PDB entries for proteins with ligands in PDBbind v2020, 1,330 entries were discarded by the filters and 2,452 entries were discarded because they were unable to pass the structure fixing and refinement modules. Our final optimized PDBbind dataset contains 15,661 unique PDB IDs and 27,757 protein-ligand complexes structures. In addition, considering that the original PDBbind general set was further filtered to create a "refined" and "core" set based on structure quality, binding data quality, and redundancy reduction\cite{pdbbind_2016}, the optimized PDBbind data yields totals of 4,969 and 279 entries in the refined and core sets, respectively. Finally, the associated binding affinity data is reorganized into a machine-readable format (csv) with comments as to the form of the experimental binding free energy data: $K_d$, $K_i$, $\mathrm{IC}_{50}$ and $\mathrm{EC_{50}}$.\\   

\begin{figure}[H]
\begin{center}
\includegraphics[width=1.0\textwidth]{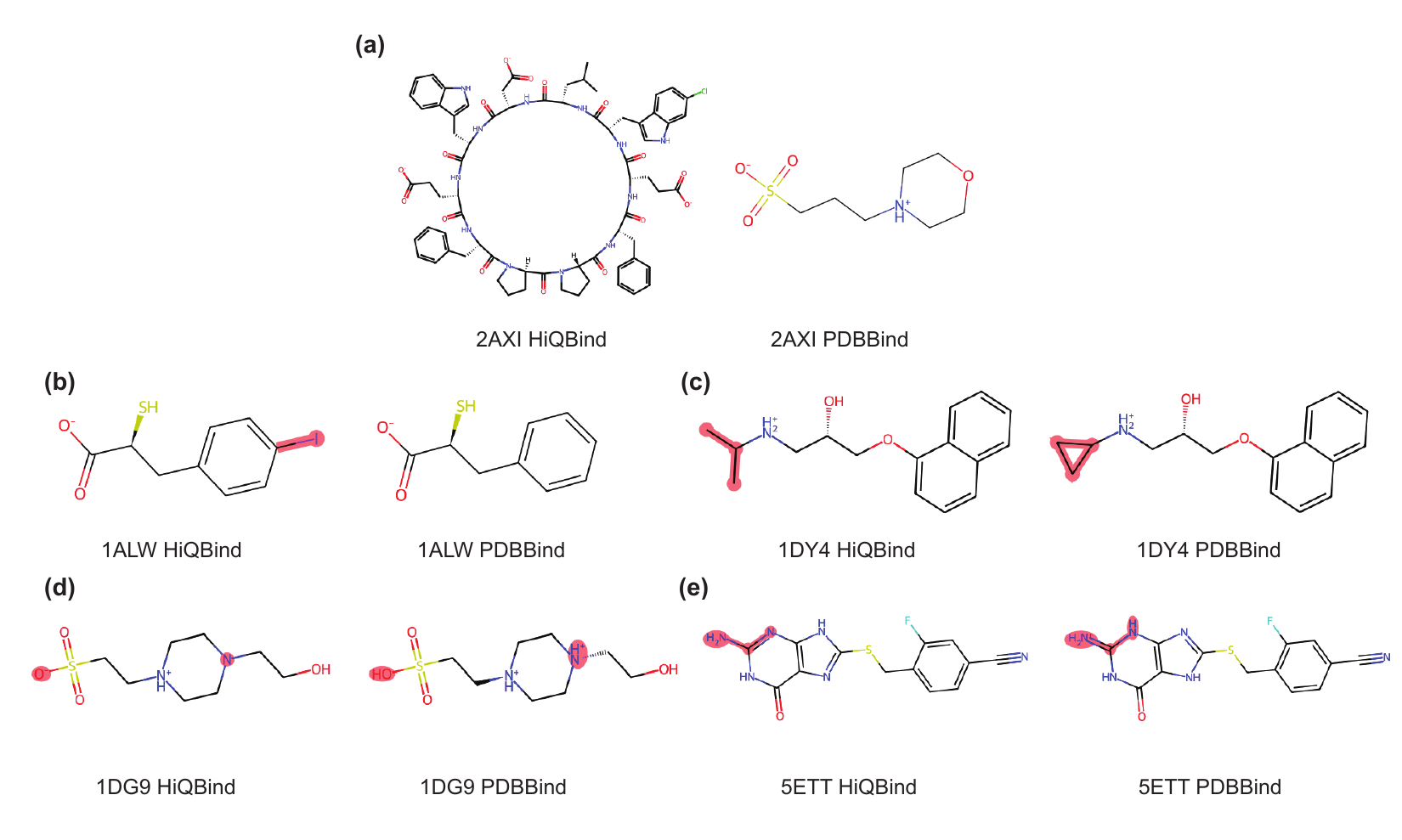}
\end{center}
\caption{\textit{Examples of corrected ligands by HiQBind-WF compared to the original PDBbind.} (a) Wrong ligand entity reported. (b) Missing atoms. (c) Wrong bond connectivity. (d-e) Undesired protonation and tautomer states. The mol2 format ligand files in PDBbind database were used for analysis.}
\label{fig:err_ligand}
\end{figure}

\noindent
\textbf{Example of refined ligand structures using HiQBind-WF.} Here, we also provide examples of the fixed ligand structures obtained from HiQBind-WF and compared with the deposited ligands in the original PDBbind dataset as provided in Figure \ref{fig:err_ligand}. In some cases, we find that some of the ligands in PDBbind are different from what was actually reported in the literature from which they were derived. For 2AXI\cite{2axi}, the ligand of interest should be the cycflic peptide-like inhibitor, not the sulfonic acid buffer. In other cases, the PDBbind ligand structures are incomplete or the bonding is incorrect (Figure \ref{fig:err_ligand}a). For example, the ligand in 1ALW\cite{1alw} is missing an iodine atom and in 1DY4\cite{1dy4} the isopropyl is falsely reported as a cyclopropyl (Figure \ref{fig:err_ligand}b,c). This type of problem may arise for historical reasons, i.e. some structures in PDBbind were derived from older version of RCSB PDB that contained these incorrect structures. We also find that HiQBind-WF yields ligands with better protonation/tautomer states. Two examples are 1DG9\cite{1dg9} and 5ETT\cite{5ett}, for which PDBbind shows that the former case contains a neutral sulfonic acid and a divalent piperazine cation motif while the latter case falsely makes a guanosine-like compound positively charged (Figure \ref{fig:err_ligand}d,e). In this case, the fixed ligand structures are more chemically feasible and also in line with the protonation states predicted by ChemAxon Marvin\cite{chemaxon}. 

HiQBind-WF also fixes a small but practical problem in PDBbind.   PDBbind provides two file formats for the ligand structure, mol2 and sdf. However, among all 19,443 entries, 45 mol2 files and 3175 sdf files cannot be processed by RDKit\cite{rdkit} (version 2024.03.4), a widely-used open-source cheminformatics tool. This may be due to the fact that these files are prepared by some other software and their sdf specifications are not compatible with RDKit. Examples are undesired aromaticity specification (oxygens tagged as aromatic to represent equivalent atoms in $\mathrm{RSO_3^-,RCOO^-,RPO_3^{2-}}$) or formal charge specification (nitrogen with 4 explicit valence tagged to be neutral). HiQBind-WF naturally addresses this technical problem because it uses RDKit\cite{rdkit} to process ligand structures.\\


\begin{figure}[H]
\begin{center}
\includegraphics[width=0.8\textwidth]{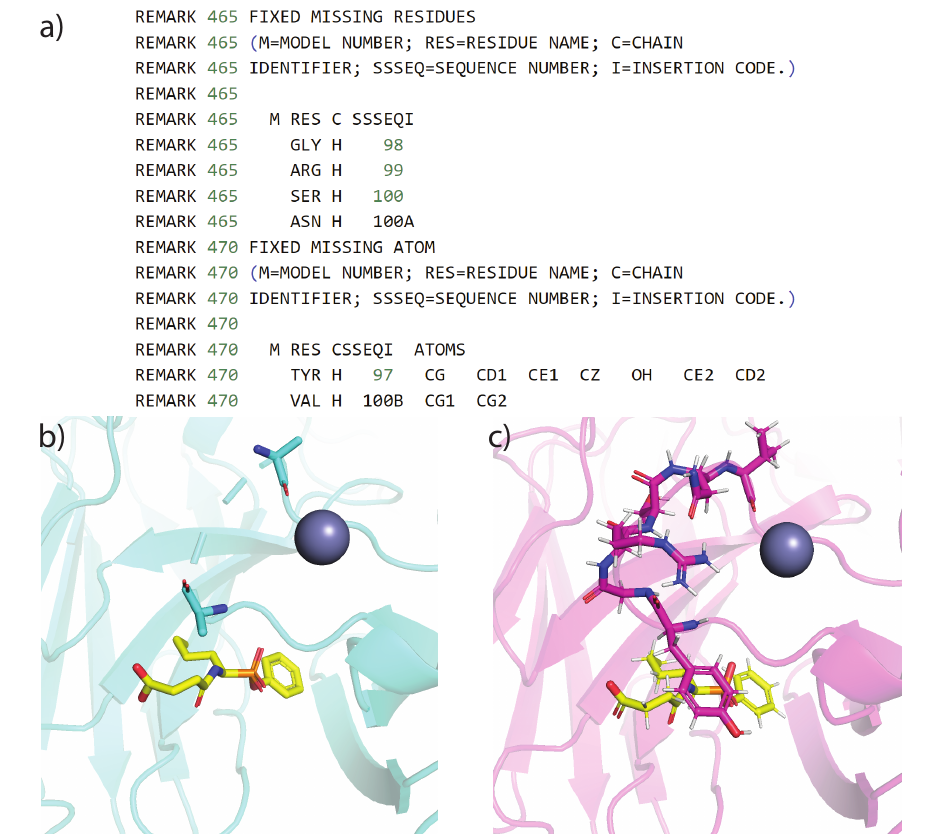}
\end{center}
\caption{\textit{Example of a refined protein structure derived from HiQBind-WF compared to the original PDBbind}. (a) The protein fixing metadata related to residue Y97 through V100B in the refined protein file. (b) Visualization of the original PDB entry with missing residues and atoms centralized at the region Y97-V100B close to the binding site. (c) Visualization of the refined protein structure after the protein-fixing module.}
\label{fig:protein-fix}
\end{figure}

\noindent
\textbf{Example of refined protein structures using HiQBind-WF.} With the protein-fixing module, users interested in training 3D-based SFs and capturing local protein-ligand interactions would benefit from a more complete protein and binding site representation. To demonstrate our protein-fixing module, Figure \ref{fig:protein-fix} shows an example of protein 1A0Q\cite{1a0q} that has both missing atoms and missing residues around the binding site. Here, the protein-fixing module first identified those missing data and fixed them based on the sequence information provided in the mmcif file header and the predefined residue templates. The reason behind using information from the mmcif header rather than the pdb "SEQRES" field is that in some of the deposited structures, missing residues are also omitted in the “SEQRES" field. As a result, an unphysical peptide bond will be placed between the start and the end of a short sequence of the missing residues, which will cause problems in training SFs. The metadata from this fixing call is stored in the refined pdb file in case users want to label the original crystal residues and repair residues differently.

\subsection{Creation of the HiQBind Dataset}
\noindent
In order to further demonstrate the utility of HiQBind-WF, we have created a new dataset of high-quality, non-covalent protein-ligand complex structures and their associated binding affinity values. To prepare the HiQBind dataset, we used two biologically relevant protein-ligand datasets as a starting point: BioLiP2\cite{biolip2} and Binding MOAD\cite{moad2023}. We downloaded the txt-formatted BioLiP database from its official website and csv-formatted dataset MOAD from its Github repository and entries with at least one reported binding affinity ($K_i$, $K_d$, $\mathrm{IC_{50}}$ or $\mathrm{EC_{50}}$) data were selected. BioLiP2 itself provides a sizable collection of protein-ligand entries deposited in RCSB PDB and enriched with multiple annotations, including binding affinity data from various sources including Binding MOAD\cite{moad2023}, BindingDB\cite{bindingdb1,bindingdb2}, and manual annotation. Although BioLiP2 encompasses much of Binding MOAD, we still found additional entries from Binding MOAD that we also include in our new dataset. Over this entire merged set seven PDB entries (2BXG, 2I30, 3T74, 3T8G, 4H57, 6TMN and 7JWN) were discarded because their binding affinities are invalid (with $K_i>10^3$ M), and all entries as part of the publicly available PDBbind v2020 dataset are not included. In total, 20,349 unique PDB entries with reported binding affinities were obtained. 

We then applied HiQBind-WF to process all these starting entries, yielding 18,160 unique PDB entries. For the 2,189 entries that failed to pass HiQBind-WF, 761 of them are discarded by the filters and the remaining 1,428 entries are due to the failure of structure fixing and refinement modules (Table S4). A large portion of the discarded datapoints are "polymers" for which it is hard to verify their structural correctness because of the difficulty in obtaining a reliable reference SMILES string which is more suitable to small molecules. Almost certainly, human inspection and expertise will rescue some of the discarded data, but the design goal here is to automate the corrections with a high throughput procedure as much as possible. 

\begin{figure}[H]
\begin{center}
\includegraphics[width=0.99\textwidth]{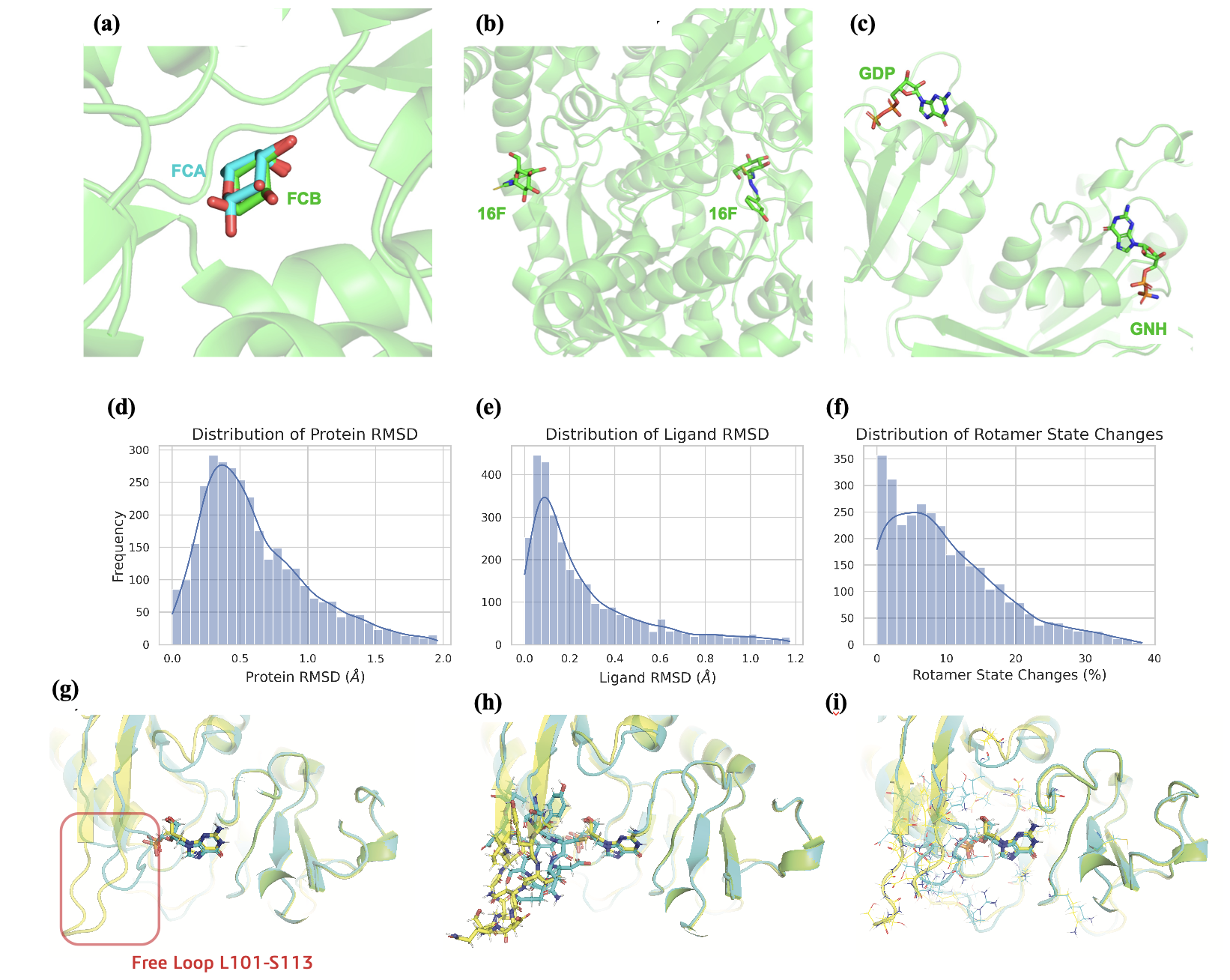}
\end{center}
\caption{\textit{Structural variations and more than one protein-ligand complexes in the same PDB entry.} (a) Two stereoisomers (FCA and FCB) contribute to the binding for 1ABF. (b) The same ligand (16F) binds to two different pockets in 3MRV(c) Two different ligands (GDP) and (GNH) bind to the protein in different pockets for 1A4R. (d) Distribution of protein RMSD. (e) Distribution of ligand RMSD. (f) Percentage of rotamer state changes for residues around the ligand binding sites. Bottom row: visualization of changes in rotamer states between chain A (blue) and chain B (yellow) of PDB ID: 3GEP. (g) Structure overlay between two chains and their respective ligands. (h) Structural differences around the free loop regions between two chains visualized as sticks. (i) Rotamer comparisons of all 29 residues that change their states across chains.}
\label{fig:multiple_ligand}
\end{figure}

At the same time we retain 32,275 protein-ligand complex structures. The reasons behind the increase in the amount of data compared to PDBbind is that we have included multiple protein-ligand complexes from the same RCSB PDB entry because: (1) multiple conformers or stereoisomers can contribute to the binding (Figure \ref{fig:multiple_ligand}a); (2) the same ligand can bind to different protein pockets (Figure \ref{fig:multiple_ligand}b); (3) there is more than one ligand with a reported binding affinity (Figure \ref{fig:multiple_ligand}c); 131 PDB entries are of this reason; and (4) for structures containing homo-multimers, structural fluctuations between chains that share sequence identity are non-negligible.  

A moderate amount of PDB entries contain multiple records of the same ligand of interest in the deposited structures. The reason lies in the fact that proteins can form various quaternary structures using copies of the same chain, and ligands as binders can interact with the macromolecule at the tertiary or quaternary level. For example, when a ligand binds to a specific chain of a homodimer protein, two PDB entries are present. PDBbind \cite{pdbbind_2016} usually keeps only one randomly-selected sample of the interacting protein-ligand complex. However, since different chains in PDB are resolved separately using their electron density maps, there are still some level of non-negligible structural variations among different copies, making them valuable data sources for training SFs. 

Although the overall protein and ligand RMSD distributions between identical chains of the same entry do not show a great difference (Figure \ref{fig:multiple_ligand}(d,e)), there is a significant amount of side-chain rotamer state changes observed for different chains as shown in Figure \ref{fig:multiple_ligand}(f). Here, following the common practice in the field\cite{rotamer_lib}, we used the angle cutoff of $60^\circ$ to any of the four side-chain torsion angles to define a switch in the rotamer states. To illustrate, the protein chains A and B for PDB entry 3GEP\cite{protein_3gep} in Figure \ref{fig:multiple_ligand}(g-h) shows that 29 out of 57 residues near the binding site have a change in their side-chain rotameric states, including 12 residues in the free loop area (L101, S103-I113). In particular, the distance between the side chain of D107 and the ligand in chain A (blue) is smaller than 4\AA, compared to chain B where the free loop is further from the ligand. Therefore, including multiple records of protein-ligand interactions with the same PDB entry can be informative and beneficial.

\begin{figure}[H]
\begin{center}
\includegraphics[width=1.0\textwidth]{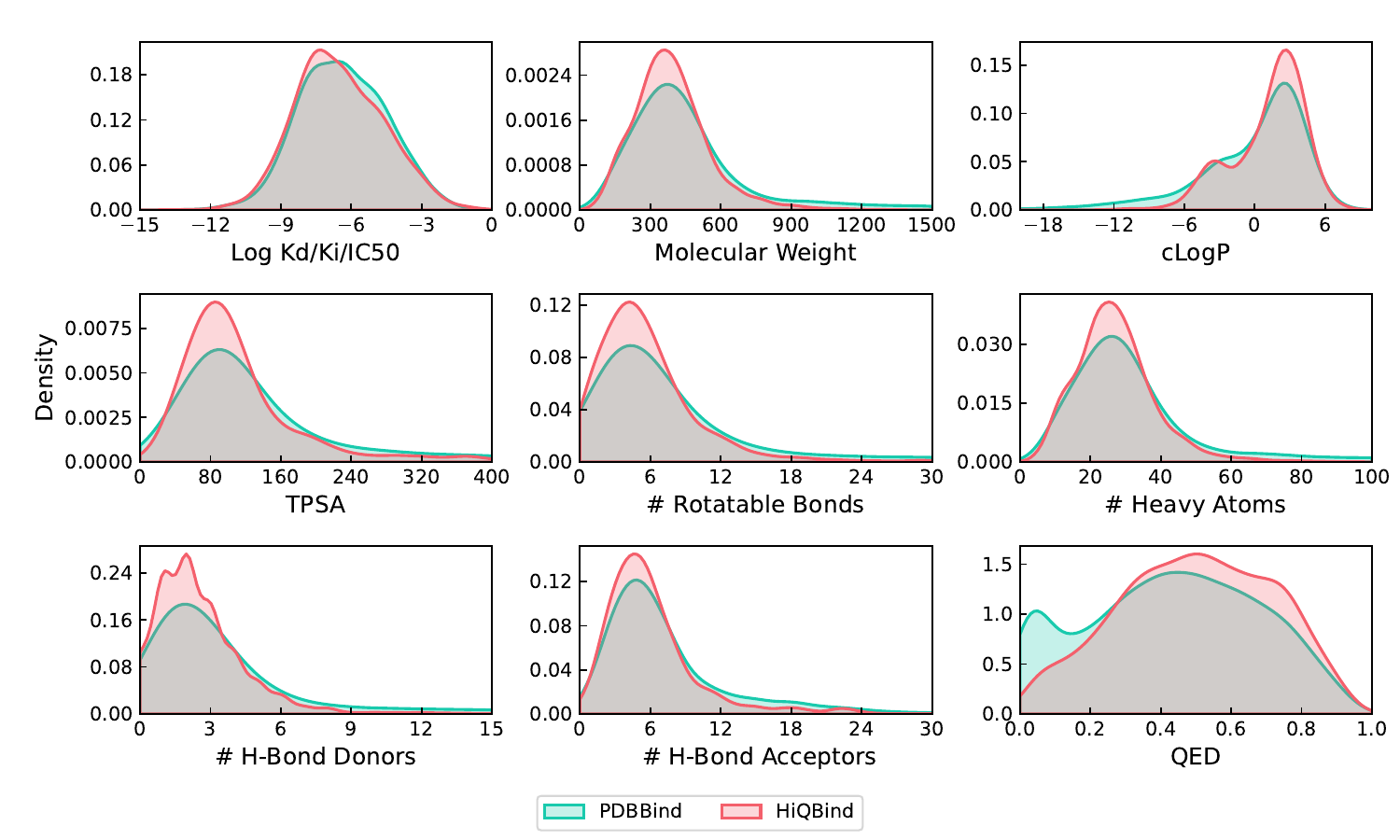}
\end{center}
\caption{\textit{Comparison of the distribution of drug-related properties of ligands and their binding affinities between PDBbind and HiQBind.} (a) Binding $K_d$, $K_i$, and $\mathrm{IC}_{50}$ values in log units. (b) molecular weight, (c) computed $\log P$ value, (d) topological polar surface area (TPSA), (e) the number of rotatable bonds, (f) the number of heavy atoms, (g) the number of hydrogen-bond donor atoms, (h)the number of hydrogen-bond acceptor atoms, and (i) quantitative estimation of drug-likeness (QED) values. }
\label{fig:dist}
\end{figure}

To characterize and validate the HiQBind dataset, Figure \ref{fig:dist} provides the distributions of binding affinities and drug-like properties compared to the PDBbind dataset. It is seen that the new HiQBind dataset shares a very close set of distributions of drug-like properties with PDBbind, especially for the binding affinities in which both datasets cover a large window of approximately 10 log units. We also noticed that HiQBind dataset is a bit more druglike (QED score) with smaller ligands having fewer rotatable bonds and better cLogP/hydrogen-bonding properties. This demonstrates the feasibility of the new HiQBind dataset as a useful resource for future SFs development, benchmarks and other structure-based drug design studies. It is also important to note that, since no standardized method exists for further filtering and data splitting, it is up to the users decide how to perform these additional operations. For example, one might filter out NMR structures and entries with $\mathrm{IC_{50}}$ or $\mathrm{EC_{50}}$ values, as done in the PDBbind refined set\cite{pdbbind_2014_1}, or split the data based on time\cite{durairaj2024plinder,pdbbindblind}, sequence similarity\cite{lppdbbind}, ECOD classifications\cite{dockgen,durairaj2024plinder}, or protein–ligand interaction profiles\cite{durairaj2024plinder}. In the Supporting Information, we also provided analysis upon the time distribution of PDB entries in HiQBind as well as its overlapping with PDBbind v2020 (Figure S1). This shall benefit users who want to do time-based splitting or create independent test set for those models that have already been trained on PDBbind v2020.

\section{CONCLUSIONS}
Many physics-based and machine-learned scoring functions used to predict protein-ligand binding affinities rely on powerful databases such as PDBbind\cite{pdbbind_2004, pdbbind_2005, pdbbind_2009, pdbbind_2014_1, pdbbind_2014_2, pdbbind2014, pdbbind_2016}, BioLiP2\cite{biolip,biolip2}, Binding MOAD\cite{moad2023}, and related databases such as Binding DB\cite{bindingdb1,bindingdb2}, Plinder\cite{durairaj2024plinder}, and Dockgen\cite{dockgen}. While central to the biomolecular and drug discovery communities, all data curation efforts require ongoing quality-control efforts. In fact, the latest PDBbind version hosted on the PDBbind+ website have performed some corrections,
but PDBbind data curated after v2021 has been commercialized such that it is only accessible to paid users, and there is no published literature describing their workflow. Hence, we are unable to make a fair comparison between the quality of their generated data compared to that  presented here. Therefore, we believe there is a need for the community to have a free open-source tool to curate high-quality protein-ligand structures in a reproducible way.

We have developed an optimization workflow, HiQBind-WF that aims to improve the structural integrity in a semi-automated way and produce high-quality structures with binding affinity annotations. 
We compared PDBbind v2020 to the structures processed with HiQBind-WF.  Differences between the complexes highlighted the strength of our workflow in assigning correct bond orders, protonation states, and protein structure refinements.
We also used this workflow to prepare HiQBind, a newly compiled dataset based on BioLiP2\cite{biolip2} and Binding MOAD\cite{moad2023} that offers high-quality, non-covalent protein-ligand complexes with binding-affinity data. HiQBind provides more than 30,000 protein-ligand structures spanning over 18,000 unique PDB entries and is feasible to be deployed in the development of scoring functions or force fields or related activities.

As an open source effort, we believe that HiQBind-WF provides a sustainable framework for continuously updating and refining protein-ligand binding datasets for drug discovery, by meeting scientific goals of  ensuring transparency and reproducibility. We also envision that structure-based drug design studies can benefit from the new HiQBind data that has no overlap with PDBbind, and thus reporting evaluation metrics upon this new dataset that will become a common practice as part of future computational modeling efforts.

\section{SUPPLEMENTARY INFORMATION}
Details of the data formats and usage notes are provided in the Supplementary Information.

\section{DATA AND CODE AVAILABILITY}
All the codes for HiQBind-WF workflow and HiQBind dataset creation are provided in a public accessible GitHub repository: https://github.com/THGLab/HiQBind under MIT License. The associated DOI is: https://doi.org/10.5281/zenodo.14903380.

The HiQBind dataset, including the protein-ligand structures, metadata information (binding affinity annotations, release year, resolution, ligand name, protein name, protein UniProt ID and various ligand properties) is publicly avaiable in Figshare: 

https://doi.org/10.6084/m9.figshare.27430305

\section{AUTHOR CONTRIBUTIONS}
Y.W., K.S., J.L. and T.H.-G. conceived the scientific direction for HiQBind workflow and HiQBind dataset and wrote the manuscript. Y.W. and K.S. wrote the codes and prepared the datasets. All authors provided comments on the results and manuscript.

We thank Michael Gilson for helpful discussions. This work was supported by National Institute of Allergy and Infectious Disease grant U19-AI171954. This research used computational resources of the National Energy Research Scientific Computing, a DOE Office of Science User Facility supported by the Office of Science of the U.S. Department of Energy under Contract No. DE-AC02-05CH11231. 


\bibliography{references}
\bibliographystyle{naturemag}
\end{document}


\title
{\textbf{Supporting Information: A Workflow to Create a High-Quality Protein-Ligand Binding Dataset for Training, Validation, and Prediction Tasks}}

\author{Yingze Wang$^{1,\Delta}$, Kunyang Sun$^{1,\Delta}$, Jie Li$^{1}$,
Xingyi Guan$^{1}$, Oufan Zhang$^{1}$, Dorian Bagni$^{1}$, 
Yang Zhang$^{4-6}$, Heather A. Carlson$^{7}$, Teresa Head-Gordon*$^{1-3}$}
 \date{}
\maketitle
\begin{center}
\vspace{-10mm}
$^1$Kenneth S. Pitzer Theory Center and Department of Chemistry, $^2$Department of Bioengineering, $^3$Department of Chemical and Biomolecular Engineering, University of California, Berkeley, CA, 94720 USA

$^4$Department of Computer Science, School of Computing, National University of Singapore, 117417,  $^5$Cancer Science Institute of Singapore, National University of Singapore, 117599, $^6$Department of Biochemistry, Yong Loo Lin School of Medicine, National University of Singapore, 117596, Singapore

$^7$Odyssey Therapeutics Inc. 1350 Highland Dr., Ann Arbor, MI, 48108, USA

$^{\Delta}$authors contributed equally

corresponding author: thg@berkeley.edu
\end{center}


\noindent

\section{Data Usage}
\noindent The HiQBind dataset can be found in the following Figshare repository:

\begin{itemize}
    \item https://doi.org/10.6084/m9.figshare.27430305
\end{itemize}

\noindent which contains the following files:

\begin{itemize}
    \item \texttt{hiqbind.tar.gz}: Protein-Ligand complexes structures in HiQBind
    \item \texttt{hiqbind\_metadata.csv}: Metadata for HiQBind dataset
    \item \texttt{hiqbind\_sm\_metadata.csv}: Metadata for subset of HiQBind in which ligands all small molecules
    \item \texttt{hiqbind\_poly\_metadata.csv}: Metadata for subset of HiQBind in which ligands are polyemers
    \item \texttt{README.md}: Description of the structural dataset and columns in the metadata csv file
\end{itemize}

The metadata files are also provided in the Github repository:

\begin{itemize}
    \item https://github.com/THGLab/HiQBind/blob/main/figshare
\end{itemize}

Unzipping the data tarball with command \texttt{tar -xzvf hiqbind.tar.gz} will yield two directories \texttt{raw\_data\_hiq\_sm} and \texttt{raw\_data\_hiq\_poly} corresponding to to the "small molecule" and "polymer" subset of HiQBind, respectively. In each directory, you will see a file structure like this: 

\begin{verbatim}
-- 1a69/
   |-- 1a69_FMB_A_240/
       |-- 1a69_FMB_A_240_ligand.pdb
       |-- 1a69_FMB_A_240_protein.pdb
       |-- 1a69_FMB_A_240_protein_hetatm.pdb
       |-- 1a69_FMB_A_240_hetatm.pdb
       |-- 1a69_FMB_A_240_ligand_refined.sdf
       |-- 1a69_FMB_A_240_protein_refined.pdb
   |-- 1a4m_FMB_B_240/
   |-- 1a4m_FMB_C_240/
-- 1a85/
\end{verbatim}

An overview of the file contents is in Table \ref{tab:file} and here is a description of the naming conventions of the files:

\begin{itemize}
    \item \texttt{1a69}: 4-letter PDB ID
    \item \texttt{FMB}: Name of the ligand. If the ligand is a polymer, it will be format like "ACE-DIP", where "ACE" is the name of the first residue and "DIP" is the name of the last residue.
    \item \texttt{A}: Ligand chain ID.
    \item \texttt{240}: Ligand residue number. If the ligand is a polymer, it will be format like "1-3", where "1" is the residue number of the first residue and "3" is the number of the last residue. Note the residue number may contain insertion code, or be a negative integer or zero.
\end{itemize}

\begin{table}[H]
    \centering
    \caption{Overview of the files in HiQBind.}
    \label{tab:file}
    \begin{tabular}{p{5cm}|p{10cm}}
         \hline
         \textbf{File} & \textbf{Description} \\
         \hline
         *\_ligand\_refined.sdf & Refined ligand structures (hydrogen added, correct bond order, better tautomer states/protonataion states) with HiQBind-WF workflow \\
         *\_protein\_refined.pdb & Refined protein structures (hydrogen added, missing atoms/residues added) with HiQBind-WF workflow \\
         *\_ligand.pdb & Ligand structure extracted from the original PDB (not processed) \\
         *\_protein.pdb & Protein structure extracted from the original PDB (not processed). A protein is defined as chains within 10 angstrom of the ligand structure. \\
         *\_hetatm.pdb & Additives' structure extracted from the original PDB (not processed) \\
         *\_protein\_hetatm.pdb & Protein structure with additives (solvents, ions) extracted from the original PDB (not processed). Additives are specified with "HETATM" atoms that are within 4 angstroms of the protein chains. \\
         \hline
    \end{tabular}
\end{table}

A description of all fields in the csv-formatted metatdata file in in Table \ref{tab:metadata}.

\begin{table}[H]
    \centering
    \begin{threeparttable}
    \caption{Overview of the HiQBind metadata fields.}
    \label{tab:metadata}
    {\small
    \begin{tabular}{l|l|p{7.5cm}}
         \hline
         \textbf{Field} & \textbf{Type} & \textbf{Description} \\
         \hline
         PDBID & string & Four-letter PDB code\\
         Resolution & string / float & Resolution of the structure or "NMR" \\
         Year & int & Initial release year in RCSB PDB database\\
        Ligand Name & string & Name of the ligand.\tnote{\textit{a}} \\
        Ligand Chain & string & Chain ID of the ligand.\\
        Ligand Residue Number & string & Residue number of the ligand.\tnote{\textit{b}} \\
        Binding Affinity Measurement & string & Type of binding affinity assay: "kd", "ki", "ic50" or "ec50".\tnote{\textit{c}} \\
        Binding Affinity Sign & string & Sign of the binding affinity measurement: "=", ">=", "<=" or "$\sim$".\\
        Binding Affinity Value & float &  Value of the binding affinity\\
        Binding Affinity Unit & string &  Unit of the binding affinity: "fM", "pM", "nM", "uM", "mM" and "M".\\
        Log Binding Affinity & float &  Binding affinity in log unit.\\
        Binding Affinity Source & string &  Source of the binding affinity annotations: "BindingMOAD", "BindingDB" or "BioLiP"\\
        Binding Affinity Annotation & string &  The annotation in the original source.\\
        Protein UniProtID & string &  UniProtID of the proteins, seperated by a comma if the ligand bound to more than one chain.\\
        Protein UniProtName & string &  Name of the proteins, separated by a comma if the ligand bound to more than one chain.\\
        Ligand SMILES & string &  SMILES of the ligand.\\
        Ligand MW & float &  Molecular weight of the ligand.\\
        Ligand LogP & float &  LogP value of the ligand computed by RDKit.\\
        Ligand TPSA & float &  TPSA value of the ligand computed by RDKit.\\
        Ligand NumRotBond & int &  Number of rotatable bonds in the ligand.\\
        Ligand NumHeavyAtoms & int &  Number of heavy atoms in the ligand.\\
        Ligand NumHDon & int &  Number of hydrogen bond donors in the ligand.\\
        Ligand NumHAcc & int &  Number of hydrogen bond acceptors in the ligand.\\
        Ligand QED & float &  QED value of the ligand computed by RDKit.\\
         \hline
    \end{tabular}
    \begin{tablenotes}
        \footnotesize
        \item[\textit{a}] If the ligand is a polymer, its name will be format like "ACE-DIP", where "ACE" is the name of the first residue and "DIP" is the name of the last residue.
        \item[\textit{b}] If the ligand is a polymer, its residue number will be format like "1-3", where "1" is the number of the first residue and "3" is the number of the last one. Note the residue number may contain insertion code, or be a negative integer or zero.
        \item[\textit{c}] In some sources where binding data is labeled as $K_a$ or $K_b$, they are converted to $K_d$ using $K_a = 1/K_d$.
    \end{tablenotes}
    }
    \end{threeparttable}
\end{table}

\begin{figure}[htbp]
  \centering
  \begin{minipage}[t]{0.6\linewidth}
    \centering
    \caption*{\textbf{(a)}}
    \includegraphics[height=6cm]{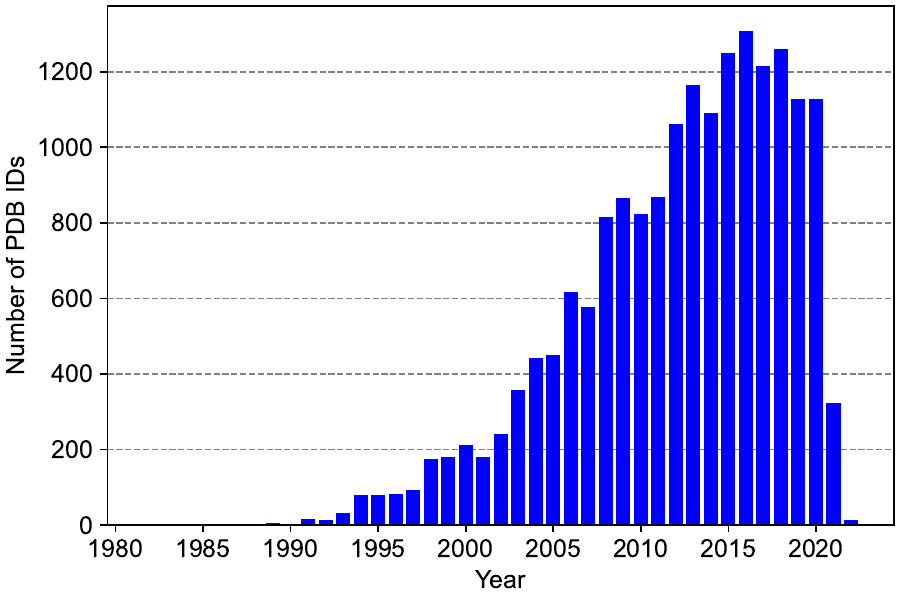}
    \label{fig:year}
  \end{minipage}
  \begin{minipage}[t]{0.3\linewidth}
    \centering
    \caption*{\textbf{(b)}}
    \includegraphics[height=5.5cm]{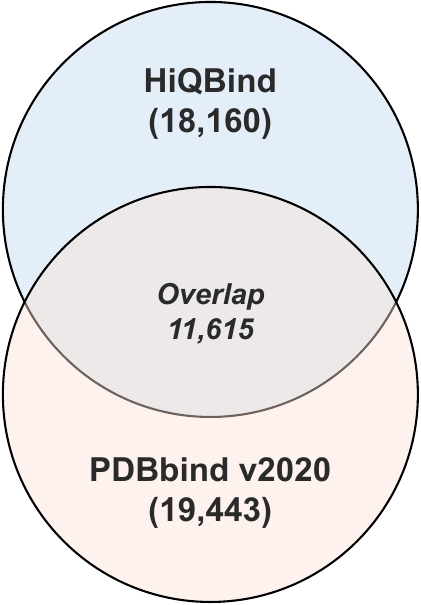}
    \label{fig:overlap}
  \end{minipage}
  \caption{\textbf{(a)} The year distribution of PDB entries in HiQBind. HiQBind contains 1,463 PDB entries deposited after 1/1/2020.  \textbf{(b)} HiQBind PDB entries overlapping with those in PDBbind v2020. Among 18,160 unique PDB IDs in HiQBind, 11,615 of them appear in PDBbind v2020.}
  \label{fig:s-analysis}
\end{figure}

Since users may want to create their own data splits based on time or create subset that does not overlap with exisiting dataset, such as PDBbind v2020, here we provide some relevant analysis in Figure \ref{fig:s-analysis}. All entries in PDBbind v2020 are before 1/1/2020, and HiQBind contains 1,463 PDB entries deposited after this date. Overall, 11,615 PDB entries out of HiQBind's total 18,160 entries are in PDBbind v2020. This overlapping is due to the fact that many entries in also appear in Binding MOAD and BindingDB, which are datasets that HiQBind queries binding-affiniy annotation from. 

\newpage

\section{Failed cases in processing HiQBind}

\begin{table}[htbp]
    \centering
    \begin{threeparttable}
    \caption{Overview of the number of entries during filter and structure fixing process of HiQBind.}
    \label{tab:hiqbind}
    \begin{tabular}{p{6cm}p{3.5cm}p{2cm}p{3cm}}
    \hline
        & \multicolumn{3}{c}{Number of unique PDB IDs} \\ 
        & Small Molecule & Polymer & Total\\
    \hline
         Before processing& 19105 & 1250 & 20349\tnote{\textit{a}}  \\
         Successfully processed\tnote{\textit{d}}& 17725\ (31572) & 438\ (703) & 18160\tnote{\textit{b}}\ \ (32275)\\
         Failed & 1380 & 812 & 2189\tnote{\textit{c}}\\
         Covalent binders & 262 & 16 & 278\\
         Steric clashes & 105 & 7 & 112 \\
         Contain uncommon elements & 355 & 6 & 361 \\
         Small ligands & 10 & 0 & 10\\
         Fail to fix structures & 638 & 783 & 1428\tnote{\textit{c}}\\
    \hline
    \end{tabular}
    \begin{tablenotes}
      \footnotesize
      \item[\textit{a}] 6 PDB entries are included in both small molecule set and polymer set: 1ga8, 1gwv, 1gx4, 1o7o, 1o9f, 4d4u.
      \item[\textit{b}] 3 PDB entries are sucessfully processed in both small molecule set and polymer set: 1gwv, 1o7o, 1o9f.
      \item[\textit{c}] 1ga8 and 4d4u are failed in both small molecule set and polymer set; 1gx4 is successful in small molecule set but failed in polymer set.
      \item[\textit{d}] In parenthesis is the number of protein-ligand structrues.
    \end{tablenotes}
    \end{threeparttable}
\end{table}

\noindent\textbf{278 entries are identified with covalent binders:}

{\small 1au0, 1ayu, 1ayv, 1ayw, 1b0f, 1b4e, 1b6a, 1c3b, 1dru, 1drw, 1erm, 1eta, 1etb, 1fsw, 1fsy, 1ga9, 1gfw, 1h2y, 1hbj, 1i72, 1iau, 1kds, 1kdw, 1ke0, 1ke3, 1lhc, 1lhd, 1ms0, 1mt5, 1my8, 1nl6, 1nlj, 1o2t, 1o41, 1o45, 1o4d, 1o4e, 1o4i, 1o5f, 1o5w, 1pi4, 1pi5, 1q6k, 1qfs, 1qhr, 1qm5, 1re1, 1rhm, 1rhr, 1rhu, 1rwm, 1rww, 1s2y, 1s3b, 1s3e, 1snk, 1sri, 1tu6, 1u9v, 1u9w, 1u9x, 1uks, 1w31, 1wof, 1yt7, 2abe, 2alv, 2amd, 2aux, 2auz, 2bxr, 2bxs, 2c72, 2c73, 2c75, 2c76, 2clx, 2dc6, 2dc8, 2dc9, 2dca, 2dcc, 2dcd, 2fj0, 2fs9, 2fzi, 2g5p, 2g5t, 2g63, 2gsu, 2hpp, 2hpq, 2hwp, 2j5f, 2jai, 2jdh, 2jdk, 2op9, 2qky, 2ql5, 2ql7, 2ql9, 2qlb, 2qlq, 2qq7, 2v6n, 2wn2, 2y4a, 2y55, 2yld, 2zu4, 2zu5, 2zz3, 2zz4, 2zz6, 3afk, 3b0r, 3bls, 3dwq, 3dz5, 3fkv, 3g3d, 3g3m, 3gzn, 3hj0, 3i06, 3i4a, 3ika, 3jwe, 3k7f, 3k83, 3k84, 3kjq, 3kw9, 3kwb, 3kwz, 3lok, 3m3c, 3m3e, 3o1g, 3ovx, 3qug, 3rdh, 3sgu, 3sw6, 3u1i, 3w2t, 3zim, 3zs0, 3zs1, 3zvt, 3zvw, 4an0, 4b16, 4bxn, 4d0l, 4dmx, 4e3n, 4hbp, 4hjs, 4hpc, 4i24, 4ll0, 4lqm, 4lrm, 4lv0, 4oyq, 4rus, 4twy, 4wkt, 4x21, 4yqm, 4ywb, 4z16, 4zzm, 4zzo, 5cyi, 5f2e, 5fct, 5feq, 5gmp, 5i3k, 5j87, 5j9y, 5j9z, 5lf3, 5lf7, 5mxq, 5u4f, 5weu, 5xdk, 5xdl, 5xxd, 5xxg, 5y9t, 5yu9, 5za2, 6alj, 6avi, 6ax1, 6b1e, 6bl1, 6c1i, 6cge, 6cha, 6di9, 6e6e, 6ezp, 6hhf, 6hhi, 6iuo, 6j99, 6jpj, 6jwl, 6jx0, 6jx4, 6jxt, 6k1s, 6lw2, 6lze, 6m0k, 6md1, 6mzw, 6o0e, 6o0f, 6oni, 6p69, 6p7i, 6pdz, 6pnm, 6pnn, 6pno, 6qw7, 6r4v, 6rnu, 6swm, 6t5y, 6tfp, 6tpm, 6v6k, 6v6o, 6v9c, 6vh4, 6vim, 6why, 6wi0, 6wtj, 6wtt, 6wxz, 6xd3, 6xhm, 6xmk, 6xr3, 6y2f, 6yi8, 6yzr, 6zpi, 7b3e, 7bwd, 7c6u, 7c8u, 7cb7, 7cbt, 7d1m, 7dpp, 7jkv, 7jxl, 7jxp, 7jxw, 7k0e, 7k1h, 7lel, 7ltn, 7mlf, 7orf, 7rbz, 7rc0, 7tob}

\noindent\textbf{112 entries are identified to exhibit steric clashes:}

{\small 1bb0, 1ca8, 1e55, 1fbp, 1h00, 1qga, 1rdy, 1vyr, 1w6p, 1ykl, 1ykp, 1z8a, 2c01, 2ime, 2j9m, 2jjk, 2r24, 2r9m, 2vj1, 2w68, 2w92, 2wnj, 2xuc, 2zyk, 3a23, 3cwd, 3d04, 3fck, 3kfy, 3lb4, 3mdj, 3n8y, 3pch, 3qce, 3qcf, 3rho, 3rv7, 3uo9, 3w4k, 3wzu, 3zrc, 4bii, 4d0m, 4e26, 4ele, 4fh7, 4fm5, 4g5y, 4gs9, 4he9, 4ic0, 4jlm, 4otj, 4ucf, 4v04, 4xi3, 4yx9, 4zx6, 4zzw, 5afk, 5dx3, 5dxb, 5ewa, 5fb7, 5fry, 5glt, 5glz, 5gm0, 5h5o, 5hjg, 5hk9, 5hn7, 5hn9, 5hyr, 5i3b, 5is0, 5ivt, 5sye, 5syf, 5t6p, 5vn1, 5wbf, 5xg4, 5z75, 6a6k, 6ah4, 6ah5, 6c7g, 6ch5, 6co5, 6cup, 6eum, 6gno, 6jz0, 6kiu, 6kiv, 6kix, 6kiz, 6n92, 6n94, 6n96, 6ncf, 6ohu, 6oo8, 6ppl, 6pww, 6qdf, 6ula, 6whv, 6wv1, 7c7h, 7cr4}

\noindent\textbf{361 entries are identified contain ligand with uncommon elements:}

{\small 1cp6, 1cws, 1d3v, 1dd7, 1dzj, 1el7, 1el8, 1esz, 1hq5, 1hyv, 1hyz, 1k2v, 1lvk, 1mu9, 1nop, 1rff, 1rfi, 1rg1, 1rg2, 1rgt, 1rgu, 1rh0, 1suo, 1svk, 1tqv, 1tqw, 1v97, 1w0g, 1wva, 1xuf, 1xuj, 1y3g, 1ztz, 2aeb, 2ato, 2bdm, 2bt3, 2c2s, 2c2t, 2cfd, 2ci0, 2cib, 2d0t, 2dy5, 2fb3, 2fdu, 2fdv, 2fdw, 2fou, 2fov, 2foy, 2gj8, 2h7r, 2ij7, 2jjp, 2jld, 2nz5, 2oi4, 2oro, 2orp, 2p8o, 2pll, 2rar, 2rav, 2rb5, 2rbk, 2v0m, 2v96, 2v97, 2w09, 2w0b, 2wfg, 2wh8, 2whf, 2wpu, 2wuz, 2wv2, 2wx2, 2xfh, 2yak, 2ydm, 3b6h, 3b99, 3ben, 3bwf, 3e6k, 3e6v, 3e9b, 3fxz, 3fy0, 3g5n, 3gph, 3i7z, 3i80, 3i8w, 3ixb, 3ixg, 3jt4, 3jt9, 3jus, 3khm, 3koh, 3ksw, 3lc4, 3mdt, 3mke, 3mmr, 3mnu, 3nxu, 3o0j, 3ozu, 3ozv, 3ozw, 3qoa, 3r9c, 3ruk, 3swz, 3t3q, 3t3r, 3t3s, 3t3z, 3tmz, 3vjs, 3vjt, 3wax, 3way, 3zg2, 3zg3, 4c0c, 4c4r, 4c4s, 4coh, 4d2y, 4d2z, 4d30, 4d31, 4d32, 4d33, 4d34, 4d35, 4d36, 4d37, 4d38, 4d39, 4d3a, 4d3b, 4d75, 4d78, 4d7d, 4dtw, 4dtz, 4du2, 4dub, 4due, 4duf, 4ehr, 4ejh, 4eji, 4enh, 4fdh, 4fil, 4g0c, 4g44, 4g45, 4h4d, 4h4e, 4hpd, 4hww, 4hxq, 4hze, 4i06, 4i4g, 4i60, 4ie2, 4ie3, 4iu0, 4iu4, 4ixu, 4ixv, 4j14, 4jfv, 4jfw, 4jjf, 4jjg, 4k9t, 4k9v, 4k9w, 4k9x, 4mdg, 4mdl, 4mdm, 4nkv, 4ob0, 4pf7, 4pk5, 4pk6, 4q3q, 4q3r, 4q3s, 4u72, 4u74, 4uch, 4uhi, 4uhl, 4uqh, 4uzi, 4v3u, 4v3v, 4v3w, 4v3x, 4v3y, 4v3z, 4wmz, 4xrz, 4xud, 4z0l, 4zdy, 4zdz, 4ze0, 4ze1, 4ze2, 4ze3, 4zee, 4zgx, 5agp, 5brv, 5c8i, 5cg5, 5cg6, 5cjh, 5e58, 5e7v, 5ek2, 5ek3, 5ek4, 5etw, 5fom, 5fqb, 5fqc, 5fsa, 5hs1, 5ibe, 5ibf, 5ibg, 5irq, 5irv, 5jqt, 5jws, 5jwu, 5jww, 5k1l, 5k7k, 5l2r, 5l92, 5l94, 5mon, 5moo, 5tz1, 5u48, 5u4a, 5u4c, 5u4e, 5vc0, 5vce, 5veu, 5wmu, 5wn8, 6a16, 6a17, 6abk, 6bbs, 6bc9, 6bcc, 6c6n, 6c8x, 6ceh, 6chi, 6cir, 6ciz, 6cr2, 6d1l, 6d1m, 6da3, 6da5, 6daa, 6dab, 6dac, 6dag, 6daj, 6e40, 6e41, 6f0a, 6ffk, 6frj, 6gcy, 6h1l, 6h1t, 6h3q, 6hwz, 6hx5, 6i96, 6j8r, 6jn6, 6kof, 6kps, 6kvq, 6kw7, 6m7x, 6ma6, 6ma7, 6msn, 6mso, 6o3i, 6pgx, 6pht, 6pu7, 6q2t, 6q37, 6q39, 6qfu, 6qfv, 6qfw, 6qfx, 6qou, 6rpn, 6rvf, 6rvk, 6rvl, 6rw1, 6sp7, 6t7u, 6t9z, 6u30, 6u7o, 6ung, 6unh, 6uni, 6unk, 6unl, 6v7c, 6v7d, 6v7e, 6v7f, 6vbn, 6wpl, 6wr0, 6wr1, 6xz8, 6xz9, 6yzq, 6yzs, 6yzt, 6yzu, 6yzv, 6yzw, 6yzx, 6zjc, 7e0p, 7e7f, 7m8v, 9nse}

\noindent\textbf{10 entries are identified to contain small ligands:}

{\small 1lg6, 1ray, 1ugb, 1ugf, 1zdq, 2ca2, 2hds, 4e5q, 4qef, 6pdv}

\noindent\textbf{1428 entries failed the structure fixing:}

{\small 11gs, 1a07, 1a2c, 1a5v, 1ad8, 1agm, 1apv, 1apw, 1aqc, 1at5, 1at6, 1axw, 1b11, 1b40, 1b5q, 1b6j, 1bdq, 1bfn, 1bm2, 1bzh, 1c0p, 1c5o, 1c5p, 1c5z, 1cka, 1ckb, 1ckp, 1clu, 1d6s, 1deh, 1dmb, 1e4w, 1e5j, 1eb1, 1ec9, 1eef, 1epq, 1evh, 1eyn, 1ez9, 1f5k, 1f86, 1fh7, 1fh8, 1fh9, 1fhd, 1fls, 1fm1, 1fwu, 1fwv, 1g3c, 1g42, 1g9r, 1ga8, 1gah, 1gai, 1gbn, 1gni, 1gnm, 1gnn, 1gno, 1gny, 1gse, 1gu3, 1gui, 1gvu, 1gvx, 1gwl, 1gwq, 1gwr, 1gww, 1h24, 1h2t, 1h2u, 1h9g, 1hew, 1hkk, 1hpb, 1hps, 1htb, 1htg, 1hut, 1i30, 1i7c, 1i7m, 1igj, 1is0, 1it6, 1j1a, 1j84, 1j8v, 1jd5, 1jfh, 1jlx, 1jn2, 1ju9, 1jvp, 1k1y, 1k7t, 1k7u, 1kjl, 1kjr, 1kl3, 1kl5, 1l6m, 1l7z, 1lax, 1lek, 1lf9, 1lkk, 1lqe, 1lr8, 1lzb, 1lze, 1lzg, 1m7d, 1m7i, 1md2, 1mf4, 1mfa, 1mfd, 1mfl, 1mhw, 1n92, 1nm5, 1nzy, 1oau, 1obx, 1oby, 1ocq, 1of4, 1ogg, 1oh3, 1oh4, 1ols, 1olu, 1olx, 1oxn, 1oxq, 1p2g, 1ph0, 1phw, 1pig, 1pl0, 1pmh, 1ppi, 1pzi, 1pzk, 1q0y, 1q4k, 1qaw, 1qiw, 1qj6, 1qj7, 1qja, 1qjb, 1qlu, 1qsc, 1ra1, 1ra9, 1rbm, 1rbz, 1rc0, 1rcv, 1rd9, 1rdj, 1rdl, 1rdm, 1rdp, 1rf2, 1rr8, 1rrj, 1rsu, 1rx9, 1s7y, 1sem, 1seu, 1shd, 1skx, 1sld, 1sle, 1slg, 1sln, 1ssq, 1str, 1sts, 1syn, 1t7d, 1t8i, 1tb4, 1tc1, 1tf8, 1tlc, 1tps, 1tsd, 1tw6, 1u5b, 1ua7, 1ugx, 1ugy, 1ukq, 1ukt, 1uld, 1ule, 1ulg, 1umw, 1ur8, 1ur9, 1utc, 1ux7, 1uxa, 1uxb, 1uz8, 1uzn, 1v0k, 1v0l, 1v11, 1v16, 1v1m, 1vwl, 1vwn, 1w1g, 1w2h, 1w3k, 1w3l, 1w3y, 1w8f, 1w8h, 1w9d, 1w9e, 1w9o, 1w9w, 1wdn, 1wdr, 1wgc, 1wu1, 1x08, 1x09, 1x7w, 1x7x, 1x7y, 1x7z, 1x80, 1xff, 1xkx, 1xl1, 1xnk, 1xt3, 1xt8, 1xzx, 1y3n, 1y3p, 1y3y, 1y79, 1y7l, 1yf4, 1yhm, 1yp9, 1yxd, 1zfp, 1zhs, 1zkk, 1zky, 2a3y, 2ak5, 2aoc, 2aod, 2aoh, 2aou, 2ar6, 2arb, 2ay7, 2az8, 2az9, 2azb, 2azc, 2bcd, 2bel, 2bgr, 2bmz, 2c05, 2c7u, 2cb8, 2cdp, 2chn, 2cjw, 2cvq, 2dt3, 2e95, 2eax, 2er9, 2euk, 2eum, 2evl, 2ff4, 2fgu, 2fgv, 2foq, 2fos, 2gfa, 2gfj, 2ggu, 2gh9, 2h13, 2h6k, 2h6n, 2h6q, 2hah, 2hj4, 2hjb, 2hl3, 2ho2, 2i74, 2ifw, 2ii3, 2imf, 2iqd, 2itk, 2j1t, 2j6o, 2jg8, 2jjb, 2mip, 2msb, 2ntb, 2nwl, 2nwx, 2nxd, 2nxl, 2nxm, 2o9v, 2oei, 2ok0, 2oor, 2ooz, 2orv, 2p4t, 2peh, 2q5a, 2q7q, 2q8y, 2qmj, 2qt5, 2qta, 2qtz, 2r0h, 2r1w, 2r1x, 2r1y, 2r23, 2r3y, 2rca, 2ri9, 2rkm, 2rkn, 2tpi, 2uue, 2uvh, 2uvi, 2uvj, 2uyq, 2v7d, 2v83, 2v86, 2v88, 2v8l, 2vco, 2vj0, 2vl1, 2vnf, 2vnk, 2vpe, 2vqz, 2vsl, 2vt5, 2vuz, 2vzl, 2w1u, 2w3o, 2w5k, 2w5l, 2w7y, 2wa8, 2wcg, 2wd3, 2wdb, 2wev, 2wgc, 2whp, 2wk2, 2wly, 2wm0, 2wn3, 2wra, 2wt0, 2wt1, 2wt2, 2wyf, 2wyn, 2x4t, 2xd3, 2xg3, 2xg9, 2xog, 2xoi, 2xom, 2xon, 2xqq, 2xwe, 2y4s, 2y64, 2y6g, 2y6l, 2yfz, 2yg0, 2yjq, 2ylg, 2yln, 2ypp, 2z8d, 2z8e, 2z8f, 2zcr, 2zcs, 2zex, 2zey, 2zga, 2zhf, 2zhk, 2zhl, 2zhp, 2zm3, 2zno, 2zq0, 2zy1, 3a06, 3ach, 3aci, 3al3, 3avf, 3ayc, 3ayd, 3b4y, 3b9a, 3bbt, 3bho, 3bib, 3biv, 3boo, 3bpc, 3bwr, 3bzi, 3c0z, 3c3d, 3c9e, 3cbs, 3d3x, 3d6o, 3d9t, 3db3, 3dcq, 3diw, 3drg, 3dz2, 3dz4, 3dz6, 3e7a, 3ebb, 3er3, 3eys, 3f3a, 3f5j, 3f5k, 3f5o, 3f5p, 3f70, 3f81, 3f9w, 3f9y, 3fi2, 3fn0, 3g0e, 3g0f, 3g3r, 3g42, 3gjf, 3gl6, 3gss, 3gxz, 3h9f, 3hkn, 3hlo, 3hzk, 3hzy, 3i6c, 3iet, 3if7, 3iiq, 3iit, 3iqg, 3iqh, 3iqi, 3iqj, 3iqv, 3isd, 3ivv, 3jvk, 3jyr, 3jzj, 3k00, 3k45, 3k8d, 3ka2, 3kce, 3kmc, 3kyf, 3kyg, 3lbl, 3lex, 3lgs, 3ll2, 3lpl, 3lq2, 3lq4, 3lqi, 3lqj, 3lsj, 3lug, 3luh, 3lw1, 3m3o, 3m3r, 3mbp, 3muk, 3mxc, 3mxy, 3n0i, 3n7y, 3n8m, 3n8w, 3nf3, 3nkx, 3nq6, 3nw3, 3nxr, 3ny3, 3o0w, 3o0x, 3o6m, 3oea, 3oeb, 3oka, 3okp, 3old, 3ole, 3olg, 3oli, 3oy8, 3p36, 3pj1, 3q8d, 3qg6, 3qmo, 3qx3, 3qxv, 3qzs, 3r42, 3rjf, 3rme, 3rss, 3rul, 3run, 3ryb, 3sm1, 3sou, 3sow, 3sp5, 3t8v, 3tcg, 3ti4, 3tib, 3tkz, 3tl0, 3tlh, 3u4s, 3u4w, 3u51, 3ur0, 3uvm, 3uwn, 3v0w, 3vf9, 3voz, 3vp1, 3vp2, 3vtr, 3vyk, 3waw, 3wch, 3wel, 3weo, 3wol, 3wom, 3won, 3wqd, 3wqv, 3zdv, 3zkk, 3zq9, 3zst, 3zvy, 3zw2, 3zyb, 3zyh, 3zyr, 3zyx, 4a23, 4a3x, 4a45, 4a7a, 4a9u, 4aa1, 4ad2, 4ad3, 4agl, 4aif, 4ap0, 4aph, 4apr, 4asl, 4auy, 4av0, 4av5, 4avi, 4avj, 4aze, 4b4q, 4b83, 4b9z, 4bdo, 4bgx, 4bh0, 4bj0, 4bnk, 4bqr, 4btl, 4c02, 4c0r, 4c1t, 4c1u, 4c5w, 4c9w, 4cd4, 4cd5, 4cd6, 4cd8, 4cdr, 4cov, 4cow, 4coy, 4coz, 4cqm, 4cub, 4d3w, 4d44, 4d4c, 4d4d, 4d4u, 4dhl, 4dro, 4drq, 4e6c, 4egi, 4ep2, 4eqj, 4er4, 4exh, 4ezq, 4ezt, 4ezy, 4fch, 4fem, 4fgx, 4fk7, 4fut, 4g0u, 4g0w, 4g5f, 4g68, 4gah, 4gk9, 4gxk, 4gzf, 4gzw, 4h39, 4h3b, 4hcz, 4hgw, 4hoe, 4hp0, 4hpi, 4hva, 4i23, 4iaw, 4iax, 4iea, 4ifi, 4igk, 4ii9, 4ikn, 4iwd, 4j48, 4jaw, 4je8, 4jih, 4jir, 4jof, 4jog, 4joh, 4joj, 4jok, 4jor, 4jvn, 4k3l, 4k3m, 4k6y, 4k75, 4k76, 4kup, 4l6t, 4lbj, 4lbk, 4lbl, 4lbm, 4lbn, 4lbo, 4lbt, 4ljh, 4lk7, 4lkd, 4lke, 4lkf, 4lkk, 4lkm, 4lmb, 4lnp, 4lo6, 4m7j, 4m9h, 4mbp, 4mdn, 4mdr, 4mh5, 4mo4, 4mr3, 4mr5, 4mx5, 4n84, 4ngp, 4ngs, 4ngt, 4nmo, 4nmp, 4nmq, 4nmr, 4nms, 4nmt, 4nmv, 4nrt, 4nxq, 4ny3, 4o6l, 4o6w, 4oee, 4oef, 4oeg, 4ojz, 4onc, 4ou3, 4owl, 4oxy, 4oyr, 4p0a, 4p7r, 4p8v, 4p8x, 4p9v, 4p9z, 4pa0, 4pfw, 4pfy, 4phv, 4pk3, 4ptc, 4q4s, 4qhp, 4ql1, 4qlk, 4qll, 4qme, 4qvb, 4qyn, 4r5e, 4r5i, 4r6o, 4r6p, 4r6q, 4r6r, 4r6t, 4r73, 4ra5, 4rdd, 4rh5, 4rmh, 4rx0, 4rxh, 4tqm, 4tro, 4u7w, 4ua8, 4uac, 4ufb, 4up4, 4utv, 4uyw, 4v27, 4w9f, 4w9n, 4wa2, 4wey, 4wq3, 4x1r, 4x3h, 4x3r, 4x3s, 4x9r, 4x9v, 4x9w, 4xbl, 4xbn, 4xbq, 4xoq, 4xur, 4xx9, 4y32, 4y5i, 4yee, 4yef, 4yk0, 4ykj, 4ykk, 4ylz, 4ym0, 4ym2, 4ynl, 4yvv, 4yw2, 4yw8, 4yyt, 4yz5, 4yzc, 4z0u, 4z1n, 4zfz, 4zh7, 4zhc, 4zhl, 4zhm, 4zo7, 4zs9, 4zwy, 4zze, 5a3o, 5a4w, 5ab0, 5ab9, 5aci, 5acw, 5acx, 5ajb, 5aom, 5apr, 5ayf, 5azg, 5b2d, 5bmm, 5bo9, 5boo, 5bpc, 5bs8, 5bta, 5btd, 5btf, 5btr, 5btv, 5bxp, 5bxr, 5bxs, 5bxt, 5c0m, 5c1m, 5cpm, 5cqj, 5cs2, 5cvd, 5cxi, 5d6y, 5d8u, 5d9j, 5dms, 5dxg, 5e1b, 5e1d, 5e1o, 5e4t, 5e50, 5e70, 5e76, 5e7g, 5e8f, 5eel, 5eie, 5elf, 5elq, 5elz, 5em9, 5ema, 5ey9, 5f08, 5f3c, 5f3e, 5f3g, 5f3i, 5f4n, 5f5k, 5f7v, 5f90, 5fa5, 5fa6, 5fiv, 5fjx, 5fjz, 5flc, 5fpp, 5fra, 5fre, 5fu2, 5fu3, 5fvs, 5g4c, 5g5z, 5g61, 5ggo, 5ggp, 5ghv, 5glu, 5glw, 5gmv, 5gu4, 5gwk, 5gx6, 5gx7, 5gza, 5gzk, 5hct, 5hda, 5heb, 5hed, 5hes, 5hey, 5hf1, 5hfb, 5hfc, 5hff, 5hla, 5hlb, 5hld, 5hlp, 5htb, 5htc, 5hza, 5hzb, 5i2f, 5i75, 5icv, 5ifu, 5igm, 5iok, 5izf, 5j41, 5j5y, 5j8u, 5jf2, 5jf3, 5jf4, 5jf5, 5jf7, 5jf8, 5jiz, 5jq7, 5jvi, 5kcf, 5kcw, 5kd9, 5kle, 5klf, 5knj, 5kox, 5l41, 5l4l, 5l79, 5l7f, 5lb7, 5ldq, 5lkc, 5ls7, 5lub, 5lyr, 5m17, 5m1z, 5m28, 5m5d, 5m77, 5may, 5mb0, 5mb1, 5mk9, 5mka, 5mks, 5mm9, 5mnh, 5mrd, 5mtt, 5mtu, 5mwj, 5mxg, 5mxr, 5n16, 5n99, 5n9c, 5nf7, 5nfa, 5npr, 5nps, 5o22, 5o4z, 5o58, 5o7u, 5o7v, 5o7w, 5oca, 5ofl, 5ofx, 5omw, 5on2, 5on3, 5onh, 5osx, 5osy, 5q0l, 5swb, 5sz2, 5t4j, 5t54, 5t55, 5t5j, 5t5l, 5t5p, 5t78, 5t7i, 5t7t, 5t8r, 5tdb, 5tdc, 5tdd, 5tef, 5tha, 5tlm, 5tln, 5tlo, 5tlu, 5tp0, 5tpc, 5ttf, 5ttg, 5ttw, 5tyi, 5tzo, 5u06, 5u0f, 5u1q, 5udk, 5uf1, 5uf4, 5ufc, 5ur1, 5v61, 5v62, 5vdk, 5vnf, 5vrw, 5vrx, 5vry, 5vrz, 5vs1, 5vs2, 5vs3, 5vv8, 5w3y, 5w7i, 5w7x, 5wei, 5wje, 5x2a, 5xo2, 5xof, 5y5u, 5yba, 5yd3, 5yd4, 5yd5, 5ygf, 5yqw, 5ysd, 5yse, 5ysf, 5yto, 5ytu, 5yx2, 5yy4, 5zbz, 5zrf, 6a56, 6a80, 6a8n, 6apr, 6apu, 6asz, 6at0, 6ayh, 6b2c, 6b3p, 6b67, 6bn0, 6brr, 6c19, 6c5h, 6c5j, 6c5k, 6c9n, 6c9p, 6c9r, 6cb5, 6cbz, 6ccu, 6cd8, 6cdg, 6cff, 6cy7, 6d6t, 6dfr, 6dmf, 6dqn, 6dqs, 6dqv, 6dqz, 6dr0, 6drc, 6e3d, 6ecz, 6eeh, 6ek3, 6eor, 6eq1, 6eqv, 6eqw, 6eqx, 6f4p, 6f4r, 6f4t, 6f57, 6f5u, 6f6i, 6f6n, 6f6s, 6f8n, 6fau, 6fav, 6faw, 6fbw, 6fby, 6ffl, 6fft, 6fhu, 6fi4, 6fi5, 6fiv, 6fjj, 6flg, 6fmn, 6fn9, 6fpu, 6fqr, 6fu1, 6fuv, 6g15, 6g9b, 6g9i, 6gd0, 6gdo, 6gex, 6gtx, 6gty, 6gtz, 6gu0, 6gw1, 6gx6, 6h0h, 6h4o, 6h4p, 6h4r, 6h4s, 6h4u, 6h4v, 6h4w, 6h4x, 6h4y, 6h50, 6h51, 6h52, 6h5w, 6hcu, 6hhp, 6hi4, 6hi6, 6hi8, 6hm0, 6hmt, 6how, 6hpg, 6hqx, 6hro, 6hs4, 6i0l, 6i1u, 6i41, 6i5p, 6i5v, 6i5w, 6i68, 6i7a, 6ial, 6iam, 6idg, 6im4, 6inz, 6jad, 6jao, 6jba, 6jue, 6kgo, 6kgq, 6kxa, 6kxb, 6l1f, 6lcf, 6ld3, 6ld4, 6ld5, 6lf2, 6lfj, 6lj2, 6lk4, 6lra, 6lrd, 6m6p, 6m8p, 6mle, 6mlo, 6n19, 6n93, 6n97, 6nlk, 6nll, 6nln, 6nxz, 6ny0, 6o5i, 6o6n, 6orv, 6p02, 6pa7, 6phx, 6pre, 6prg, 6pwv, 6q9w, 6qdx, 6qpl, 6qsw, 6qsx, 6qux, 6r18, 6r3m, 6r6v, 6r6w, 6rav, 6rd2, 6rfh, 6rk4, 6rt6, 6rt7, 6rti, 6ryo, 6s1i, 6sey, 6sgf, 6slo, 6spw, 6spx, 6sq0, 6swn, 6swo, 6swq, 6swx, 6t1d, 6t7z, 6tg4, 6th7, 6tkm, 6tmp, 6tmq, 6tmz, 6tn0, 6tn2, 6tpk, 6tpx, 6tpy, 6tpz, 6tv4, 6twu, 6twx, 6u4q, 6u4t, 6u5m, 6u67, 6u8p, 6u8v, 6u8w, 6u8x, 6u90, 6u91, 6uc3, 6udc, 6v2e, 6vhz, 6vuf, 6vw7, 6wf5, 6wj7, 6wm1, 6wo2, 6wp5, 6wy7, 6x23, 6x3v, 6x9c, 6x9i, 6x9j, 6xwd, 6y2g, 6y58, 6yb4, 6yqn, 6yqo, 6yqr, 6yqs, 6yt1, 6yt2, 6z4q, 6z4s, 6z84, 6zb0, 6zb1, 6zb2, 6zb3, 6zfm, 6zin, 6zpd, 7a5m, 7at8, 7bvt, 7c67, 7c68, 7c69, 7c6g, 7c6i, 7c6k, 7c6m, 7c6t, 7c6w, 7c70, 7jjc, 7k0l, 7k0q, 7k5e, 7k5f, 7k5g, 7k5h, 7kkp, 7kme, 7lan, 7luf, 7n7x, 7otm, 7oty, 7pw4, 7pw5, 7pw6, 7pw7, 8icj, 8ico, 8icw, 8icx, 8icy, 8kme, 9icd, 9icu

\textbf{Note}: 1gx4 failed only in small moecule set.
}

\section{Failed cases in processing PDBBind}

\begin{table}[htbp]
    \centering
    \begin{threeparttable}
    \caption{Overview of the number of entries during filter and structure fixing process of PDBBind.}
    \label{tab:pdbbind}
    \begin{tabular}{p{6cm}p{3.5cm}p{2cm}p{3cm}}
    \hline
        & \multicolumn{3}{c}{Number of unique PDB IDs} \\ 
        & Small Molecule & Polymer & Total\\
    \hline
         Before processing& 16738 & 2705 & 19443  \\
         Successfully processed\tnote{\textit{a}}& 14915\ (26626) & 746\ (1131) & 15661\ (27757)\\
         Failed & 1823 & 1959 & 3782\\
         Covalent binders & 857 & 101 & 958\\
         Steric clashes & 128 & 37 & 165 \\
         Contain uncommon elements & 197 & 9 & 206 \\
         Small ligands & 1 & 0 & 1\\
         Fail to fix structures & 640 & 1812 & 2452\tnote{\textit{c}}\\
    \hline
    \end{tabular}
    \begin{tablenotes}
      \footnotesize
      \item[\textit{a}] In parenthesis is the number of protein-ligand structrues.
    \end{tablenotes}
    \end{threeparttable}
\end{table}

\noindent\textbf{958 entries are identified with covalent binders:}

{\small 1a09, 1a46, 1a5g, 1a61, 1ad8, 1aht, 1amn, 1au0, 1au2, 1auj, 1avp, 1awf, 1awh, 1ayu, 1ayv, 1b0f, 1b5g, 1bgo, 1bio, 1bjr, 1bmq, 1bwn, 1c3b, 1doj, 1e34, 1e37, 1eas, 1eat, 1ekb, 1ero, 1erq, 1exw, 1f1j, 1f7b, 1f9e, 1fsw, 1fsy, 1ft4, 1g37, 1ga9, 1gbt, 1gfw, 1ggd, 1gzg, 1gzv, 1h0w, 1h1b, 1h8y, 1hbj, 1i72, 1i8j, 1iau, 1iem, 1iew, 1inc, 1jlx, 1k2i, 1kds, 1kdw, 1ke0, 1ke3, 1l6s, 1l6y, 1lhc, 1lhd, 1lhe, 1lhf, 1lhg, 1llb, 1me3, 1me4, 1mem, 1mns, 1mpl, 1ms0, 1ms6, 1mwt, 1mxo, 1my8, 1nc6, 1njt, 1nju, 1nkm, 1nl6, 1nlj, 1nms, 1no9, 1npz, 1nqc, 1ny0, 1nyy, 1o2t, 1o41, 1o43, 1o45, 1o4a, 1o4d, 1o4e, 1o4i, 1o4k, 1o5f, 1ong, 1onh, 1p01, 1p02, 1p03, 1p04, 1p05, 1p06, 1p10, 1pau, 1pi4, 1pi5, 1q6k, 1qcp, 1qfs, 1qhr, 1qj1, 1qj6, 1qj7, 1qtn, 1qwu, 1qx1, 1re1, 1rhj, 1rhk, 1rhm, 1rhq, 1rhr, 1rhu, 1rtl, 1rww, 1rwx, 1rxp, 1snk, 1sre, 1sri, 1tbz, 1tlo, 1tmb, 1tu6, 1tyn, 1u9v, 1u9w, 1u9x, 1ukt, 1uod, 1vgc, 1vsn, 1w10, 1w12, 1w14, 1w31, 1x6u, 1y19, 1yk7, 1ylv, 1yly, 1ym1, 1yms, 1yt7, 1z6f, 1zom, 1zpb, 1zpc, 2a4g, 2a4q, 2ajb, 2ajd, 2ajl, 2alv, 2asu, 2aux, 2auz, 2bdl, 2bz5, 2clx, 2eep, 2f9u, 2f9v, 2fda, 2fj0, 2fm2, 2fs8, 2fs9, 2ftd, 2fxr, 2g5p, 2g5t, 2g63, 2g83, 2gbf, 2gbg, 2gph, 2gsu, 2gvf, 2h5d, 2h5i, 2h5j, 2h65, 2ha0, 2hds, 2hob, 2hwo, 2hwp, 2i03, 2i72, 2jal, 2jbv, 2jdh, 2k0x, 2lpr, 2mlm, 2nqg, 2nqi, 2o7e, 2o7v, 2o9a, 2obo, 2oc0, 2oc1, 2oc7, 2op3, 2op9, 2oz2, 2q3z, 2q80, 2q9m, 2q9n, 2qaf, 2qcn, 2qky, 2ql5, 2ql7, 2ql9, 2qlb, 2qlf, 2qlj, 2qlq, 2qnz, 2qq7, 2qve, 2r4b, 2r6n, 2rcx, 2rjr, 2rjs, 2uzj, 2v6n, 2vgc, 2wap, 2wgi, 2wig, 2wij, 2wik, 2wj1, 2wj2, 2woq, 2wzx, 2wzz, 2xcn, 2xdm, 2xdw, 2xe4, 2xk1, 2xlc, 2xln, 2xni, 2xow, 2xu1, 2xu3, 2xu4, 2xu5, 2xzc, 2y2h, 2y2i, 2y2j, 2y2k, 2y2n, 2y2p, 2y4a, 2y55, 2y59, 2yj2, 2yj8, 2yj9, 2yjb, 2yjc, 2z3z, 2zu3, 2zu4, 2zu5, 2zz6, 3a73, 3b1t, 3b1u, 3b9s, 3bar, 3bh3, 3bjm, 3bls, 3blt, 3blu, 3bm6, 3bm8, 3bwk, 3c9e, 3d4f, 3d62, 3dz5, 3e90, 3ewu, 3ex3, 3ex6, 3eyd, 3fkv, 3fmq, 3fmr, 3fnm, 3g3d, 3g3m, 3gjs, 3gpj, 3gzn, 3h0e, 3hd3, 3hha, 3hj0, 3hwn, 3i06, 3i4a, 3ibc, 3ika, 3iut, 3k7f, 3k83, 3k84, 3kjf, 3kjn, 3kjq, 3kqa, 3kw9, 3kwb, 3kwz, 3lj7, 3lle, 3lok, 3lox, 3lpr, 3lxs, 3m3c, 3m3e, 3mbz, 3mkf, 3mxr, 3mxs, 3n4c, 3n5e, 3ns7, 3nzi, 3o1g, 3o6t, 3o86, 3o87, 3o88, 3of8, 3oj8, 3oli, 3opp, 3opr, 3ovx, 3oyp, 3p8e, 3pa8, 3pcb, 3pcf, 3pch, 3pr0, 3qkv, 3qsd, 3rdh, 3rjm, 3s1y, 3s22, 3s3q, 3s3r, 3sji, 3sjo, 3sn8, 3sna, 3snb, 3snc, 3snd, 3sv6, 3sv7, 3sv8, 3svv, 3sz9, 3szb, 3t9t, 3tdz, 3tjm, 3u1i, 3ufa, 3ur9, 3v4j, 3v4x, 3v6r, 3v6s, 3vb4, 3vb5, 3vb6, 3vb7, 3vgc, 3w2p, 3w2q, 3w2t, 3wnr, 3wns, 3wnt, 3zcz, 3zeb, 3zim, 3zmh, 3zmi, 3zmj, 3zot, 3zs0, 3zs1, 3zvt, 3zvw, 4amx, 4amy, 4amz, 4an0, 4an1, 4axm, 4bcb, 4bcc, 4bs5, 4bsq, 4bxn, 4ccd, 4d8e, 4d8i, 4dcd, 4dkt, 4dmy, 4e3i, 4e3j, 4e3k, 4e3l, 4e3m, 4e3n, 4e3o, 4ede, 4eej, 4efg, 4ehm, 4ejf, 4est, 4exz, 4f49, 4fgt, 4fzc, 4fzg, 4gd6, 4ght, 4gk7, 4gkc, 4gs6, 4hbp, 4hcu, 4hcv, 4hnp, 4hrc, 4hrd, 4i7c, 4i7d, 4i9o, 4i9r, 4i9s, 4imq, 4imz, 4inh, 4inr, 4int, 4inu, 4ivk, 4j5p, 4j70, 4jg6, 4jg7, 4jg8, 4jj7, 4jj8, 4jje, 4jmx, 4jr0, 4kqo, 4kw6, 4l0l, 4len, 4lv1, 4lv2, 4lv3, 4lys, 4m1j, 4m8t, 4mao, 4mbf, 4mnv, 4mvn, 4mz4, 4mzo, 4mzs, 4nk3, 4nnn, 4nnw, 4no1, 4no6, 4no8, 4no9, 4o7d, 4ob2, 4ool, 4oon, 4osf, 4pid, 4piq, 4pis, 4pji, 4pkb, 4pl3, 4pl4, 4pl5, 4pnc, 4q1s, 4q2k, 4qbb, 4qkx, 4qps, 4qq5, 4qqc, 4qvl, 4qvm, 4qvn, 4qvp, 4qvq, 4qvv, 4qvw, 4qvy, 4qw0, 4qw1, 4qw3, 4qw4, 4qw5, 4qw6, 4qw7, 4qwf, 4qwg, 4qwi, 4qwj, 4qwk, 4qwl, 4qwr, 4qws, 4qwu, 4qwx, 4qxj, 4qz0, 4qz1, 4qz2, 4qz3, 4qz4, 4qz5, 4qz6, 4qz7, 4qzw, 4qzx, 4r02, 4r17, 4r18, 4r3b, 4r6v, 4rsp, 4ruu, 4s2i, 4tkn, 4tky, 4twy, 4u0g, 4u0x, 4uuq, 4vgc, 4wbg, 4wks, 4wkt, 4wku, 4wkv, 4wm9, 4wmc, 4wsj, 4wsk, 4wx4, 4wx6, 4wx7, 4wyy, 4wz4, 4wz5, 4x0u, 4x21, 4x68, 4x69, 4x6j, 4xbb, 4xbd, 4xcu, 4xjr, 4xuz, 4yas, 4yec, 4yhf, 4yqm, 4yqu, 4yqv, 4yrs, 4yrt, 4yv8, 4z16, 4zro, 5acb, 5ahj, 5c1x, 5c1y, 5c20, 5c91, 5cls, 5cyi, 5d11, 5d6e, 5d6f, 5d9p, 5dg6, 5dgj, 5dp4, 5dp5, 5dp6, 5dp7, 5dp8, 5dp9, 5dpa, 5e0g, 5e0h, 5e0j, 5e7r, 5eb2, 5ee8, 5eec, 5est, 5f02, 5f90, 5fa7, 5fao, 5fap, 5faq, 5fas, 5fat, 5foo, 5fq9, 5g0q, 5gmp, 5gnk, 5gso, 5gty, 5gwa, 5gwz, 5h6v, 5hg5, 5hg7, 5hg8, 5hg9, 5hl9, 5hlb, 5hld, 5i23, 5i24, 5inh, 5j5d, 5j7p, 5j7s, 5j87, 5j8i, 5j8x, 5j9y, 5j9z, 5jh6, 5jk3, 5kre, 5kyk, 5l6h, 5l6i, 5l6j, 5l6o, 5l6p, 5lc0, 5lcj, 5lck, 5lpr, 5mae, 5maj, 5mjb, 5mqy, 5mxq, 5ne1, 5ne3, 5ngf, 5npb, 5nud, 5nwz, 5om9, 5orl, 5swh, 5sys, 5t66, 5t6f, 5t6g, 5tdi, 5teh, 5tg1, 5tg2, 5tg4, 5tg5, 5tg6, 5tg7, 5tig, 5toz, 5tts, 5ttu, 5ttv, 5tyj, 5tyk, 5tyl, 5tyn, 5tyo, 5typ, 5u4f, 5u4g, 5ug8, 5ug9, 5ugc, 5v4q, 5v88, 5vnd, 5vqe, 5vqv, 5vqx, 5vqy, 5vqz, 5w12, 5w13, 5w14, 5wac, 5wad, 5wae, 5waf, 5wag, 5wdl, 5wej, 5wfj, 5x02, 5x5g, 5x79, 5xhr, 5xyz, 5yof, 5yu9, 5za2, 5zde, 5zdg, 5zwf, 6a87, 6af9, 6afa, 6afc, 6afd, 6afe, 6aff, 6afg, 6afh, 6afi, 6afj, 6afl, 6alz, 6ary, 6ax1, 6b0v, 6b0y, 6b1e, 6b1f, 6b1h, 6b1j, 6b1o, 6b1w, 6b1x, 6b1y, 6b41, 6b95, 6bib, 6bic, 6bid, 6bkx, 6bl1, 6bl2, 6bq0, 6cha, 6cn8, 6cqt, 6cqz, 6czu, 6d3g, 6d8e, 6da4, 6db4, 6dge, 6dud, 6e5b, 6e5g, 6e5s, 6e7m, 6ert, 6euv, 6eyz, 6f34, 6f6r, 6fdq, 6fdu, 6ffn, 6ffs, 6fv1, 6fv2, 6g7f, 6g8m, 6g8n, 6g9f, 6g9s, 6gch, 6gcr, 6gop, 6gxy, 6gzy, 6h0u, 6hgy, 6hhg, 6hhh, 6hhi, 6hhj, 6htc, 6htd, 6htp, 6htr, 6hub, 6huc, 6huq, 6huu, 6huv, 6hv4, 6hv5, 6hv7, 6hva, 6hvr, 6hvs, 6hvt, 6hvu, 6hvv, 6hvw, 6i0x, 6ib0, 6ib2, 6ic5, 6ic6, 6iuo, 6iyv, 6iyw, 6j6m, 6jpj, 6k1s, 6lpr, 6m8w, 6m8y, 6m9c, 6m9d, 6m9f, 6mhb, 6mhc, 6mhd, 6mhm, 6mkq, 6mny, 6mu1, 6mzw, 6n4t, 6n9p, 6n9t, 6nd3, 6nng, 6nnr, 6nvg, 6nvh, 6nvi, 6nvj, 6nvl, 6o8i, 6oim, 6ovz, 6p8x, 6p8y, 6p8z, 6pgo, 6pgp, 6pnm, 6pnn, 6pno, 6q35, 6q5b, 6qft, 6qg4, 6qg7, 6qho, 6qhr, 6qmu, 6qw7, 6qw8, 6qw9, 6qwa, 6qwb, 6r4v, 6rjp, 6rmm, 6rn6, 6rn7, 6rn9, 6rne, 6rni, 6rnu, 6rrm, 6rtn, 6s1s, 6s9w, 6s9x, 6skb, 6skd, 6un1, 6un3, 7gch, 7lpr, 8lpr, 9lpr}

\noindent\textbf{165 entries are identified to exhibit steric clashes:}

{\small 1ba8, 1bb0, 1ca8, 1e55, 1gj5, 1gvk, 1h9l, 1ibc, 1jbd, 1jrs, 1nu8, 1oxg, 1pop, 1xxh, 1ykp, 1yyy, 1zzz, 2a4r, 2jjk, 2l75, 2p59, 2pre, 2psx, 2q6f, 2qpj, 2qqs, 2r9m, 2vcb, 2vj1, 2w68, 2w92, 2wnj, 2ww0, 2xuc, 3aav, 3atp, 3b3x, 3bcn, 3d04, 3dkj, 3e0p, 3e16, 3fck, 3fv7, 3fyz, 3fzc, 3gjq, 3gjt, 3gpe, 3gxy, 3kl8, 3lce, 3lpg, 3mo0, 3o0u, 3qce, 3qcf, 3qzq, 3rsb, 3rv7, 3sgv, 3t2c, 3tyq, 3uo9, 3wdc, 3wdd, 3wde, 3wzu, 3zmq, 3zrc, 4a2a, 4e26, 4ele, 4f3h, 4g5y, 4ga3, 4gs9, 4h38, 4h3a, 4he9, 4jlm, 4kiu, 4li5, 4lil, 4loj, 4p00, 4ps0, 4ps1, 4qlq, 4qls, 4qlt, 4qlu, 4qlv, 4v04, 4w5j, 4whs, 4ym4, 4yv2, 4yx9, 4z7f, 4z7q, 4zx6, 5afk, 5dpw, 5dx3, 5ewa, 5f5b, 5fb7, 5g1p, 5gsw, 5h5o, 5hk9, 5hn7, 5hn9, 5htb, 5hvp, 5ivt, 5izk, 5jer, 5n6s, 5ngb, 5npf, 5nqe, 5ouh, 5t6p, 5v5o, 5vfn, 5wbf, 5xg4, 5xw6, 5yjo, 5ykp, 5yr4, 5yr5, 5yr6, 5yun, 5z1d, 6a6k, 6a73, 6b22, 6byz, 6c7g, 6ccx, 6dqb, 6eqs, 6eum, 6fmp, 6hmy, 6ijl, 6ikm, 6jij, 6jki, 6jz0, 6k4r, 6kjf, 6myn, 6n55, 6n92, 6n94, 6n96, 6nt2, 6nzg, 6ohu, 6pk7, 6pvs}

\noindent\textbf{206 entries are identified to contain ligand with uncommon elements:}

{\small 1cp6, 1d3v, 1dzj, 1esz, 1hq5, 1hyv, 1hyz, 1k2v, 1lkx, 1lvk, 1nxy, 1nym, 1pq3, 1wva, 1y3g, 2aeb, 2cfd, 2cfg, 2fou, 2fov, 2foy, 2jld, 2p8o, 2pll, 2v0c, 2v96, 2wf5, 2wfg, 2wq4, 2yak, 2ydm, 2z97, 3bwf, 3c7n, 3csl, 3cst, 3e6k, 3e6v, 3e81, 3e9b, 3fxz, 3fy0, 3ixg, 3m1s, 3mg0, 3mke, 3mmr, 3mnu, 3o0j, 3p3h, 3p3j, 3p44, 3p55, 3pup, 3q4c, 3qlb, 3qo9, 3rj7, 3ro0, 3sjt, 3skk, 3sl0, 3sl1, 3vjs, 3vjt, 3w8o, 3wax, 3way, 3wc5, 3whw, 3zp9, 4aw8, 4daw, 4dcx, 4dcy, 4ehr, 4fea, 4fil, 4g0c, 4h4d, 4h4e, 4hmq, 4hww, 4hxq, 4hze, 4i06, 4i60, 4ido, 4ie2, 4ie3, 4iu0, 4iu4, 4ixu, 4ixv, 4jfv, 4jfw, 4jhq, 4jjf, 4jjg, 4k6t, 4kai, 4kb7, 4kbi, 4kii, 4l6q, 4ob0, 4ob1, 4q3q, 4q3r, 4q3s, 4rlp, 4u5t, 4xkc, 4z46, 5agi, 5agj, 5agr, 5ags, 5agt, 5dhf, 5fjw, 5fom, 5fqb, 5fqc, 5fsb, 5hj9, 5hja, 5hki, 5hlm, 5ll7, 5m29, 5m2q, 5m34, 5m3b, 5mgi, 5mnx, 5mny, 5mo2, 5mon, 5moo, 5mos, 5nxx, 5nxy, 5od5, 5tgy, 5u48, 5u4a, 5u4c, 5u4e, 5ujo, 5vrl, 5wys, 5zeq, 6abk, 6ajz, 6b30, 6bbs, 6bc9, 6c8x, 6ceh, 6d1l, 6d1m, 6e4v, 6frj, 6h3q, 6hwz, 6hx5, 6i0k, 6i0p, 6i96, 6i97, 6ibs, 6ibv, 6j8q, 6j8r, 6jn3, 6jn4, 6jn5, 6jn6, 6msn, 6mso, 6naf, 6nti, 6ntj, 6pgx, 6pht, 6q2y, 6q30, 6q37, 6q39, 6q92, 6qaf, 6qfu, 6qfv, 6qfw, 6qfx, 6rmf, 6rsa, 6rvf, 6rvk, 6rvl, 6rw1, 6skc, 6u7o, 6u7p, 6uhu}

\noindent\textbf{1 entry are identified to contain small ligands:}

{\small 6eu6}

\noindent\textbf{2452 entries failed the structure fixing:}

{\small 11gs, 1a07, 1a0t, 1a2c, 1a37, 1a3e, 1abf, 1abt, 1af2, 1agm, 1apb, 1apv, 1apw, 1aqc, 1at5, 1at6, 1atl, 1aze, 1azx, 1b11, 1b2m, 1b40, 1b6j, 1bap, 1bdl, 1bdq, 1bm2, 1bm6, 1bsk, 1bt6, 1bux, 1bzh, 1c5o, 1c5p, 1c5z, 1cka, 1ckb, 1clu, 1cpi, 1cyn, 1czq, 1d4w, 1d6s, 1d8e, 1dkd, 1dmb, 1dva, 1dxp, 1e03, 1e5j, 1eb1, 1ec9, 1eef, 1ej4, 1eoj, 1eol, 1epq, 1eub, 1evh, 1ez9, 1f47, 1f4y, 1f5k, 1ff1, 1fh7, 1fh8, 1fh9, 1fhd, 1fls, 1fwu, 1fwv, 1g42, 1g6g, 1g9r, 1ga8, 1gag, 1gah, 1gai, 1gmy, 1gni, 1gnj, 1gnm, 1gnn, 1gno, 1gny, 1gu3, 1gui, 1gvu, 1gvx, 1gwm, 1gwq, 1gwr, 1gwv, 1gzc, 1h00, 1h07, 1h24, 1h25, 1h26, 1h27, 1h28, 1h2t, 1h2u, 1h5v, 1h6e, 1hc9, 1hgt, 1hkj, 1hkk, 1hkm, 1hps, 1htg, 1i3z, 1i6v, 1i7c, 1i7m, 1i8h, 1i8i, 1idg, 1igj, 1iht, 1ikt, 1ilq, 1iq1, 1is0, 1it6, 1iwq, 1j19, 1j1a, 1j4q, 1jd5, 1jd6, 1jfh, 1jh1, 1jm4, 1jmq, 1jn2, 1jp5, 1jpl, 1juq, 1jvp, 1k1y, 1k9q, 1kat, 1kc5, 1kcs, 1kjr, 1kl3, 1kl5, 1kna, 1kne, 1l6m, 1lek, 1lf8, 1lf9, 1lkk, 1ll4, 1lqe, 1lt5, 1lxh, 1m7d, 1m7i, 1mf4, 1mfa, 1mfd, 1mhw, 1mpa, 1mv0, 1n3w, 1n4m, 1n5z, 1n7m, 1nde, 1ngw, 1nlo, 1nlp, 1nlt, 1ny2, 1o9k, 1oau, 1obx, 1ocn, 1ocq, 1od8, 1oeb, 1ogg, 1oh4, 1oj5, 1ok7, 1oko, 1ols, 1olu, 1olx, 1om9, 1orw, 1osg, 1osv, 1ov3, 1ow6, 1ow7, 1ow8, 1oxn, 1oxq, 1oy7, 1ozv, 1p28, 1p2g, 1p4u, 1pcg, 1pdq, 1ph0, 1pig, 1pl0, 1pmx, 1ppi, 1pum, 1pxh, 1py1, 1pyw, 1pzi, 1q4k, 1qaw, 1qi0, 1qiw, 1qja, 1qjb, 1qm5, 1qsc, 1r17, 1r2b, 1r6z, 1rdj, 1rdl, 1rdn, 1rgj, 1s9v, 1shd, 1sje, 1sld, 1sle, 1slg, 1sln, 1sm3, 1sps, 1ssq, 1str, 1sts, 1szm, 1t29, 1t2v, 1t37, 1t79, 1t7d, 1t7f, 1t7r, 1tc1, 1tet, 1ths, 1tl9, 1tps, 1ttv, 1tyr, 1u8t, 1uef, 1ugx, 1ugy, 1uh1, 1uj0, 1ujj, 1ujk, 1ukh, 1ule, 1ulg, 1umw, 1upk, 1ur9, 1urc, 1urg, 1utc, 1uti, 1uvu, 1ux7, 1uxa, 1uxb, 1uz8, 1v0k, 1v0l, 1v0m, 1v0n, 1v11, 1v16, 1v1m, 1vj6, 1vr1, 1vwl, 1vwn, 1w1g, 1w2h, 1w3k, 1w3l, 1w70, 1w80, 1wdn, 1wdq, 1wdr, 1ws5, 1wu1, 1x11, 1x8s, 1x9d, 1xb7, 1xff, 1xhm, 1xn2, 1xn3, 1xt3, 1xt8, 1y3a, 1y3n, 1y3p, 1y3y, 1ybg, 1ybo, 1yhm, 1ymx, 1yp9, 1yvh, 1ywi, 1yxd, 1yy6, 1z3t, 1z3v, 1zfp, 1zkk, 1zky, 1zub, 2a25, 2aez, 2aof, 2aoh, 2aoi, 2aoj, 2aou, 2aq9, 2auc, 2ay7, 2az8, 2az9, 2azb, 2azc, 2azm, 2b1q, 2b1r, 2b2v, 2b7f, 2bba, 2bcd, 2bgn, 2bgr, 2bmz, 2br8, 2byp, 2c1n, 2c9t, 2ce9, 2cht, 2ci9, 2cia, 2co0, 2d1x, 2d2v, 2df6, 2dwx, 2e7l, 2e95, 2e98, 2e9a, 2e9c, 2eh8, 2emt, 2er0, 2er9, 2euk, 2eum, 2evl, 2ez5, 2f5t, 2f6j, 2fci, 2fgu, 2fgv, 2flu, 2fr8, 2frd, 2fsa, 2fts, 2fuu, 2fx9, 2fys, 2g6q, 2gfa, 2ggu, 2gh9, 2h13, 2h2d, 2h2e, 2h2g, 2h2h, 2h6k, 2h6q, 2h9m, 2h9n, 2h9p, 2hah, 2hdx, 2hj4, 2hjb, 2hkf, 2hmh, 2hrp, 2ig0, 2igv, 2igw, 2itk, 2iv9, 2ivz, 2j7w, 2j9n, 2jb5, 2jbu, 2jdk, 2jdl, 2jg8, 2jjb, 2jk9, 2jkr, 2jkt, 2jmj, 2jnw, 2jq9, 2jqk, 2k1q, 2k3w, 2kgi, 2knh, 2kup, 2l0i, 2l65, 2l7u, 2l8j, 2lcs, 2lct, 2liq, 2llq, 2lp8, 2m0o, 2m3o, 2mg5, 2mip, 2mkr, 2mov, 2mow, 2mpa, 2mwy, 2n3k, 2n7b, 2nn8, 2nwl, 2nwn, 2nxd, 2nxl, 2nxm, 2o9k, 2o9r, 2o9v, 2odd, 2oei, 2oi9, 2ooz, 2peh, 2pem, 2pl9, 2pmc, 2pnx, 2pv3, 2q7q, 2q7y, 2q8y, 2qbx, 2qic, 2qki, 2qmj, 2qt5, 2qta, 2qtr, 2qv7, 2qwe, 2r02, 2r03, 2r05, 2r0h, 2r0y, 2r1w, 2r1x, 2r1y, 2r23, 2r2b, 2r3c, 2r3y, 2r5b, 2r7g, 2rfy, 2ri9, 2rkm, 2rkn, 2rok, 2rol, 2rvn, 2srt, 2tpi, 2uw0, 2uyq, 2uz6, 2v7d, 2v83, 2v85, 2v86, 2v87, 2v88, 2vl1, 2vnf, 2vpe, 2vpg, 2vr3, 2vsl, 2vwf, 2vxj, 2w0p, 2w0z, 2w10, 2w16, 2w2u, 2w3o, 2w47, 2w6c, 2w6t, 2w6u, 2w73, 2w76, 2w77, 2w78, 2w7y, 2w9r, 2wa8, 2wcg, 2wd3, 2whp, 2wk2, 2wly, 2wlz, 2wm0, 2wp1, 2wyf, 2wyn, 2wzf, 2x2i, 2x3t, 2x4t, 2x4z, 2x52, 2x6w, 2x6x, 2x6y, 2x85, 2xaf, 2xag, 2xah, 2xaj, 2xaq, 2xas, 2xcs, 2xct, 2xg3, 2xg9, 2xhs, 2xl2, 2xl3, 2xn6, 2xn7, 2xog, 2xoi, 2xqq, 2xrw, 2xs0, 2xs8, 2xwd, 2xwe, 2xxn, 2xzq, 2y06, 2y07, 2y1n, 2y36, 2y4m, 2y4s, 2y6s, 2y8i, 2y8o, 2y9g, 2y9q, 2ydt, 2yhw, 2yjq, 2ylc, 2yln, 2ymt, 2ynr, 2yns, 2ypp, 2yq6, 2z5o, 2z5s, 2z5t, 2zcr, 2zcs, 2zg3, 2zga, 2zgm, 2zm3, 2zpk, 2zq0, 2zy1, 2zym, 2zyn, 3afk, 3al3, 3alt, 3ap4, 3ap7, 3ary, 3arz, 3as3, 3ask, 3asl, 3au6, 3avf, 3avg, 3avh, 3ax5, 3aya, 3ayc, 3ayd, 3b3s, 3b95, 3bbb, 3bbt, 3bg8, 3bho, 3bim, 3bpc, 3btr, 3bu6, 3bu8, 3bum, 3bun, 3buo, 3buw, 3bux, 3bzi, 3c0z, 3c1n, 3c6w, 3c94, 3cbs, 3cfs, 3cfv, 3ck7, 3ck8, 3ckb, 3coj, 3cs8, 3d1e, 3d1f, 3d3x, 3d45, 3d6o, 3d9k, 3d9l, 3d9m, 3d9n, 3d9o, 3d9p, 3dab, 3dcq, 3diw, 3dla, 3dnj, 3dow, 3dpo, 3drf, 3drg, 3dri, 3ds1, 3ds9, 3dvp, 3dz2, 3dz4, 3dz6, 3e7a, 3e8u, 3ebb, 3eg6, 3ehn, 3eht, 3emh, 3eqs, 3er3, 3ery, 3eu7, 3evc, 3evd, 3evf, 3eyf, 3eys, 3eyu, 3f3a, 3f5j, 3f5k, 3f5l, 3f5p, 3f69, 3f70, 3f81, 3f9w, 3f9y, 3fbr, 3fdm, 3fdt, 3fi2, 3fn0, 3fqa, 3fuc, 3fv8, 3g0e, 3g0f, 3g2s, 3g2t, 3g2u, 3g2v, 3g2w, 3g3r, 3g42, 3g5v, 3g5y, 3g7l, 3gds, 3ggw, 3ghe, 3gl6, 3gsm, 3gss, 3gv6, 3gxz, 3h52, 3h6z, 3h91, 3h9f, 3hkn, 3hkt, 3hlo, 3hqh, 3hs8, 3hs9, 3hzk, 3hzv, 3hzy, 3i02, 3i5r, 3i6c, 3i8t, 3i90, 3i91, 3iet, 3if7, 3ifl, 3ifo, 3ifp, 3iit, 3iiw, 3iiy, 3ij0, 3ij1, 3ijy, 3ikc, 3iqg, 3iqh, 3iqi, 3iqj, 3iqq, 3iss, 3isw, 3iux, 3ivq, 3ivv, 3iw7, 3jpx, 3juq, 3jvk, 3jyr, 3jzg, 3jzh, 3jzj, 3k00, 3k26, 3k27, 3k48, 3k8d, 3ka2, 3kmc, 3krd, 3ktr, 3kyf, 3kyg, 3kze, 3l3q, 3l3x, 3l3z, 3l6x, 3lbl, 3lgl, 3lgs, 3lk1, 3lnj, 3lnz, 3lpl, 3lq2, 3lq4, 3lqi, 3lqj, 3luo, 3m3o, 3m3r, 3m53, 3m54, 3m55, 3m56, 3m57, 3m58, 3m59, 3m5a, 3mbp, 3me9, 3mea, 3met, 3meu, 3ml4, 3mp1, 3mp6, 3muk, 3mxc, 3mxy, 3n5u, 3nf3, 3nfk, 3nfl, 3nii, 3nij, 3nil, 3nin, 3nkx, 3nsn, 3nti, 3nw3, 3ny3, 3o0e, 3o1d, 3o1e, 3o6l, 3o6m, 3ob0, 3ob1, 3ob2, 3odi, 3odl, 3ogx, 3oka, 3okp, 3old, 3ole, 3olg, 3omc, 3omg, 3oq5, 3oy8, 3oyw, 3p4f, 3pdh, 3pfp, 3pgu, 3pj1, 3pkn, 3plu, 3pma, 3poa, 3pp7, 3pqz, 3psl, 3puj, 3puk, 3pxe, 3q5u, 3q6s, 3q8d, 3qfy, 3qfz, 3qg6, 3ql9, 3qlc, 3qmk, 3qn7, 3qnj, 3qo2, 3qs4, 3qxd, 3qxv, 3qzt, 3qzv, 3r42, 3r93, 3rbq, 3rdv, 3rg2, 3rl7, 3rl8, 3rme, 3rqe, 3rqf, 3rqg, 3rtx, 3rul, 3rum, 3run, 3rv6, 3rv8, 3rz9, 3rzi, 3s7f, 3shb, 3shv, 3sm1, 3so6, 3sou, 3sov, 3sow, 3stj, 3sw9, 3sxu, 3szm, 3t5i, 3t6r, 3t7g, 3t83, 3t8v, 3tcg, 3tdu, 3tf6, 3tf7, 3tg5, 3th0, 3ti4, 3tib, 3tiw, 3tkz, 3tl0, 3tlh, 3tpx, 3tsz, 3twr, 3tws, 3twu, 3twv, 3tww, 3twx, 3tzd, 3u3f, 3u78, 3ual, 3uat, 3ud7, 3ud8, 3ud9, 3uda, 3ued, 3uef, 3ueo, 3ui2, 3uig, 3uih, 3uii, 3uij, 3uik, 3upk, 3ur0, 3uri, 3uvk, 3uvl, 3uvm, 3uvn, 3uvo, 3uvu, 3uvw, 3uvx, 3uw9, 3uwl, 3ux0, 3uxg, 3uyr, 3uzd, 3v2o, 3v30, 3v3b, 3v43, 3v4t, 3v7d, 3va4, 3vf9, 3vfj, 3voz, 3vp1, 3vp2, 3vp3, 3vp4, 3vtr, 3vzg, 3w37, 3waw, 3wcb, 3wch, 3wdz, 3wp0, 3wp1, 3wqv, 3wqw, 3wsy, 3wut, 3wuu, 3wuv, 3zdv, 3zev, 3zha, 3zhf, 3zi8, 3zjt, 3zju, 3zjv, 3zke, 3zkf, 3zlv, 3zmp, 3zmt, 3zmu, 3zmv, 3zmz, 3zn0, 3zn1, 3zq9, 3zqi, 3zst, 3zvy, 3zyb, 3zyh, 3zyr, 4a0j, 4a1w, 4a23, 4a4c, 4a50, 4a7j, 4a9t, 4a9u, 4aa1, 4aa2, 4abi, 4abj, 4ad2, 4ad3, 4agl, 4aif, 4aom, 4ap0, 4aph, 4apr, 4auy, 4av0, 4av5, 4avi, 4avj, 4ay6, 4ayp, 4aze, 4b4n, 4b4q, 4b60, 4b83, 4b8o, 4b8p, 4b8y, 4b9h, 4b9w, 4b9z, 4ba3, 4bea, 4bg6, 4bgx, 4blb, 4bpi, 4bpj, 4btl, 4bv2, 4bxu, 4c0r, 4c16, 4c1t, 4c1u, 4c1w, 4c4n, 4c5w, 4c9w, 4cc2, 4cc7, 4cd4, 4cd5, 4cd6, 4cd8, 4cdr, 4ch2, 4ch8, 4ciz, 4cpq, 4cps, 4cpx, 4csy, 4cy1, 4czs, 4d1d, 4d4d, 4de7, 4dhl, 4djs, 4dma, 4dow, 4dro, 4ds1, 4dx9, 4e35, 4e3b, 4e6c, 4e81, 4e9c, 4e9d, 4edu, 4egi, 4elb, 4elg, 4elh, 4ep2, 4eqf, 4eqj, 4er4, 4erq, 4ery, 4erz, 4es0, 4esg, 4ewr, 4exh, 4ezo, 4ezq, 4ezt, 4ezy, 4f14, 4f20, 4fbx, 4fcm, 4fe9, 4fem, 4fgx, 4fgy, 4fk7, 4fmn, 4fmo, 4fmq, 4fn5, 4ft2, 4fut, 4g0a, 4g5f, 4g68, 4g69, 4gah, 4gao, 4gj8, 4glr, 4glx, 4gne, 4gnf, 4gng, 4gq6, 4gvc, 4gvd, 4gw1, 4gw5, 4gwi, 4gxl, 4gy5, 4gye, 4gzf, 4gzw, 4gzx, 4h36, 4h39, 4h3b, 4h3q, 4hcz, 4hfz, 4hgc, 4hp0, 4hpi, 4hpy, 4hs6, 4hs8, 4htp, 4hva, 4hy9, 4hyb, 4i2z, 4i31, 4i32, 4i33, 4i67, 4i7b, 4iaw, 4iax, 4ib5, 4ifi, 4igk, 4igq, 4ii9, 4ikn, 4ipn, 4is6, 4iur, 4iut, 4iuu, 4iuv, 4iwd, 4j09, 4j24, 4j26, 4j2c, 4j3u, 4j48, 4j4v, 4j73, 4j77, 4j7i, 4j84, 4j8g, 4j8r, 4j8s, 4jc1, 4jck, 4je8, 4jfx, 4jfz, 4jg0, 4jg1, 4jiz, 4jjq, 4jmg, 4jmh, 4jof, 4jog, 4joh, 4joj, 4jok, 4k0o, 4k0u, 4k3l, 4k3m, 4k63, 4k64, 4k66, 4k67, 4k6u, 4k6v, 4k6w, 4k6y, 4k72, 4k75, 4k76, 4kc1, 4kc2, 4kc4, 4kmd, 4kn7, 4kom, 4kon, 4ktu, 4kup, 4kvm, 4kx8, 4l1u, 4l58, 4l6t, 4lbl, 4lbo, 4lg6, 4ljh, 4lk6, 4lk7, 4lkd, 4lke, 4lkf, 4lkg, 4lkh, 4lkk, 4lkm, 4ln2, 4lnf, 4lno, 4lnp, 4lp6, 4lq3, 4lte, 4m1d, 4m7j, 4mbp, 4mdn, 4mdr, 4mg5, 4mo4, 4mr3, 4mr5, 4mrd, 4mx5, 4mz5, 4mz6, 4mzf, 4mzh, 4mzj, 4mzk, 4mzl, 4n3w, 4n6g, 4n7g, 4n7h, 4n7j, 4n7y, 4n84, 4nb3, 4ngn, 4ngp, 4ngq, 4ngs, 4ngt, 4nku, 4nl1, 4nmo, 4nmp, 4nmq, 4nmr, 4nms, 4nmt, 4nmv, 4nrk, 4nrl, 4nrt, 4nuf, 4nw2, 4nxq, 4ny3, 4o0r, 4o36, 4o3t, 4o3u, 4o42, 4o45, 4o4y, 4o62, 4o6w, 4oak, 4odk, 4odl, 4odm, 4odn, 4odp, 4odq, 4oee, 4oef, 4oeg, 4oel, 4oem, 4ofl, 4onf, 4oru, 4orx, 4ory, 4ou3, 4ouj, 4ov5, 4oyk, 4oz1, 4p0a, 4p0b, 4p0n, 4p4s, 4pft, 4pfu, 4pgc, 4phv, 4pl6, 4pli, 4pn1, 4pnw, 4po7, 4pry, 4psx, 4ptc, 4pvo, 4pxf, 4pz5, 4pz8, 4q1e, 4q4s, 4q6f, 4qaa, 4qc1, 4qf7, 4qfl, 4qfn, 4qfo, 4qfp, 4qh7, 4qh8, 4qhp, 4ql1, 4qlk, 4qll, 4qme, 4qq4, 4qqi, 4qsk, 4qxt, 4qy8, 4r1e, 4r3s, 4r6t, 4ra1, 4ra5, 4rh5, 4rhu, 4ris, 4rme, 4rqi, 4rqz, 4rrv, 4rxh, 4rxz, 4tk1, 4tk2, 4tk3, 4tk4, 4tmp, 4tnw, 4tt2, 4tw8, 4twt, 4tzm, 4tzn, 4tzq, 4u0a, 4u0b, 4u0c, 4u0d, 4u2w, 4u68, 4u6x, 4u7t, 4u90, 4ua8, 4uac, 4ud7, 4ue1, 4um9, 4umn, 4utn, 4utr, 4utv, 4utx, 4uu5, 4uu7, 4uu8, 4uua, 4uub, 4uw1, 4ux9, 4uxj, 4v1f, 4v27, 4w4z, 4w50, 4w5a, 4w9f, 4w9n, 4wci, 4wey, 4wht, 4why, 4wj7, 4wko, 4wph, 4wq3, 4wrq, 4wv6, 4wy7, 4wym, 4x0z, 4x13, 4x14, 4x1n, 4x1p, 4x1q, 4x1r, 4x1s, 4x34, 4x3e, 4x3h, 4x3i, 4x3k, 4x3r, 4x3s, 4x6h, 4x6s, 4x8n, 4x8p, 4x9r, 4x9v, 4x9w, 4xc2, 4xek, 4xgz, 4xh2, 4xqu, 4xtp, 4xx9, 4xxh, 4xyn, 4y32, 4y3b, 4y5i, 4yb5, 4yc8, 4ydf, 4ydn, 4yee, 4yef, 4yhp, 4yhz, 4yje, 4yjl, 4yk0, 4ykj, 4ykk, 4ym2, 4ynl, 4yoz, 4ysi, 4yw2, 4yy6, 4yyi, 4yym, 4yyn, 4yyt, 4yz5, 4yzc, 4z0d, 4z0e, 4z0f, 4z0u, 4z1n, 4z2o, 4z2p, 4z68, 4z7i, 4z83, 4z88, 4z89, 4z8m, 4zdu, 4zeb, 4zhl, 4zhm, 4znx, 4zs9, 4zwy, 5a0e, 5a2i, 5a2j, 5a2k, 5a3h, 5a3o, 5ab0, 5ab1, 5ab9, 5abp, 5acw, 5acx, 5ajc, 5ajo, 5ajp, 5aom, 5apr, 5awt, 5awu, 5ayf, 5azg, 5b2d, 5b4w, 5b56, 5b6g, 5bjt, 5bmm, 5btr, 5btv, 5c0m, 5c11, 5c13, 5c1m, 5c6v, 5c7e, 5c7f, 5cbm, 5cfa, 5cil, 5cin, 5cqj, 5cqx, 5cr7, 5cs2, 5csz, 5cvd, 5cw8, 5cxi, 5d0j, 5d1u, 5d2a, 5d6y, 5d7e, 5dah, 5dif, 5dms, 5dtj, 5duw, 5dxb, 5dxe, 5dxg, 5e0l, 5e0m, 5e1b, 5e1d, 5e1o, 5e2v, 5e2w, 5e4w, 5e8f, 5eay, 5eel, 5eeq, 5eie, 5ekg, 5elf, 5elq, 5em9, 5ema, 5emb, 5eoc, 5eok, 5epp, 5esq, 5eta, 5etf, 5etu, 5euk, 5ewz, 5ey8, 5ey9, 5eyz, 5ez0, 5f08, 5f2u, 5f3c, 5f3e, 5f3g, 5f3i, 5f4n, 5f67, 5f88, 5fb0, 5fb1, 5ff6, 5fh6, 5fiv, 5fjx, 5fkj, 5fos, 5fpi, 5fpp, 5fyq, 5g5z, 5g60, 5g61, 5g6u, 5gg4, 5ggo, 5ggp, 5ghv, 5glu, 5gmi, 5gmj, 5gmv, 5gp7, 5gs4, 5gtr, 5gu4, 5gwy, 5gx6, 5gx7, 5h1e, 5h5q, 5h5r, 5h5s, 5hct, 5hda, 5heb, 5hed, 5hes, 5hex, 5hey, 5hf1, 5hfb, 5hfc, 5hff, 5hhx, 5hjb, 5hjc, 5hjd, 5hkh, 5hlp, 5hog, 5hpm, 5htc, 5huw, 5huy, 5hyq, 5hyr, 5i25, 5i2f, 5i2i, 5i8c, 5iaw, 5ick, 5icv, 5icx, 5icy, 5icz, 5id0, 5id1, 5ifu, 5igm, 5igq, 5ijj, 5ijp, 5iok, 5iop, 5ir1, 5itf, 5iv2, 5ivn, 5ivz, 5ix1, 5ixt, 5iy4, 5iyv, 5iz6, 5izf, 5izj, 5j19, 5j31, 5j3v, 5j41, 5j5x, 5j7j, 5j8u, 5j9k, 5jek, 5jeo, 5jf2, 5jf3, 5jf4, 5jf5, 5jf7, 5jf8, 5jin, 5jiy, 5jjm, 5jlz, 5jm4, 5jop, 5jq7, 5jqb, 5jr2, 5jvi, 5jy0, 5k5c, 5k6s, 5kez, 5kgn, 5klr, 5klt, 5knj, 5kqd, 5ksu, 5ksv, 5kzp, 5l0c, 5l0h, 5l3f, 5l3g, 5l7f, 5l7k, 5lax, 5lb7, 5lbq, 5lgp, 5lgq, 5lgr, 5lgs, 5lrk, 5lsh, 5lso, 5lu2, 5lub, 5lvx, 5ly1, 5ly2, 5ly3, 5lyr, 5lzh, 5m17, 5m1z, 5m28, 5m5d, 5m63, 5m77, 5mav, 5may, 5mb1, 5mby, 5mgx, 5mk1, 5mk3, 5mk9, 5mka, 5mks, 5mlo, 5mlw, 5mm9, 5mng, 5mnh, 5mo0, 5moq, 5mrd, 5mtw, 5mwj, 5mxo, 5mxr, 5myk, 5myo, 5myx, 5n16, 5n31, 5n7b, 5n7g, 5n7x, 5n8e, 5n8j, 5n8t, 5n8w, 5n99, 5n9n, 5nfa, 5nin, 5njx, 5nne, 5npr, 5nps, 5nw8, 5nwk, 5nx2, 5nxq, 5o22, 5o45, 5o4y, 5o4z, 5o58, 5o5m, 5ocj, 5ods, 5ofx, 5ogl, 5ok6, 5osy, 5ot3, 5oua, 5ous, 5oxk, 5oxl, 5oxm, 5oxn, 5oy3, 5oyd, 5q0l, 5sve, 5svi, 5svx, 5svy, 5svz, 5swf, 5sz2, 5szb, 5szc, 5t1i, 5t1k, 5t1l, 5t1m, 5t31, 5t52, 5t54, 5t6j, 5t78, 5t7s, 5t8r, 5t90, 5tdb, 5tdr, 5tdw, 5tef, 5teg, 5th2, 5th7, 5tha, 5tkj, 5tkk, 5tln, 5tp0, 5tpb, 5tpc, 5tq1, 5tqs, 5ttf, 5ttg, 5ttw, 5twg, 5twh, 5tyi, 5tzo, 5u06, 5u0f, 5u1q, 5u2j, 5u66, 5u6k, 5ufc, 5uff, 5umz, 5un1, 5unj, 5ur1, 5uw5, 5uwi, 5uwj, 5uwp, 5v1d, 5v1y, 5v2p, 5v2q, 5v3r, 5v4b, 5v6y, 5va9, 5vb9, 5vdk, 5vk0, 5vkm, 5vlh, 5vlk, 5vll, 5vlp, 5vnb, 5vqi, 5vtb, 5vzu, 5vzy, 5w0l, 5w0q, 5w38, 5w4e, 5w5s, 5w5u, 5w6i, 5w6r, 5w6t, 5w6u, 5w7i, 5w7j, 5w7x, 5w94, 5wa1, 5wa4, 5wbk, 5wbl, 5wei, 5wg8, 5wgd, 5wgq, 5wir, 5wkl, 5wkm, 5wle, 5wqd, 5wtt, 5wxf, 5wxg, 5wxh, 5wxo, 5wxp, 5wyr, 5x72, 5xgl, 5xhs, 5xhz, 5xo2, 5xof, 5xs8, 5xup, 5xvw, 5xwr, 5xxf, 5xxk, 5xyf, 5y1u, 5y20, 5y21, 5y53, 5y59, 5y5u, 5y5w, 5y6k, 5y7w, 5y97, 5yba, 5yc1, 5yc2, 5yc3, 5yc4, 5yco, 5ygd, 5ygf, 5yjy, 5ypo, 5ypp, 5ypw, 5yqw, 5yto, 5ytu, 5yv5, 5yvx, 5yy4, 5yy9, 5yyz, 5yzd, 5z89, 5z95, 5zbz, 5zia, 5zjy, 5zjz, 5zk5, 5zk7, 5zk9, 5zml, 5znp, 5znr, 5zoo, 5zop, 5zuj, 6a30, 6a5e, 6a6w, 6a80, 6a8g, 6a8n, 6a9c, 6a9o, 6abp, 6aox, 6apr, 6apu, 6ar2, 6asz, 6at0, 6au5, 6ax4, 6axj, 6axk, 6axp, 6ayh, 6ayn, 6azk, 6azl, 6b27, 6b2c, 6b5m, 6b5o, 6b5r, 6b5t, 6b67, 6bcr, 6bcy, 6bd1, 6bgg, 6bhd, 6bhe, 6bhh, 6bhi, 6bij, 6bil, 6bin, 6bir, 6biv, 6bix, 6biy, 6biz, 6bj2, 6bmi, 6bnt, 6buu, 6bvb, 6bvh, 6bw3, 6bw4, 6byk, 6c4u, 6c5h, 6c5j, 6c5k, 6c9n, 6c9p, 6c9r, 6cb5, 6cct, 6ccu, 6cd8, 6cdg, 6cdm, 6cdo, 6cdp, 6cer, 6cf6, 6cgt, 6cjv, 6co4, 6d07, 6d08, 6d1u, 6d3x, 6d3y, 6d3z, 6d40, 6d4o, 6d6t, 6df1, 6df2, 6do5, 6drt, 6dub, 6e49, 6e5x, 6e8k, 6e8m, 6ecz, 6eeh, 6egw, 6eiz, 6ek3, 6em6, 6em7, 6ema, 6eo0, 6epy, 6epz, 6eq1, 6eqv, 6eqw, 6eqx, 6er3, 6eru, 6esa, 6evp, 6eww, 6ex0, 6ezi, 6f08, 6f09, 6f55, 6f5m, 6f5u, 6f6d, 6f6i, 6f6n, 6f6s, 6f7t, 6f8g, 6fam, 6fau, 6fav, 6faw, 6fbw, 6fby, 6fc6, 6fel, 6fhu, 6fi4, 6fi5, 6fiv, 6fkp, 6fkq, 6fky, 6fkz, 6flg, 6fmn, 6fn9, 6fpu, 6fsd, 6fse, 6fu1, 6fvn, 6fx1, 6fzf, 6fzj, 6fzp, 6g0q, 6g15, 6g2n, 6g47, 6g6x, 6g84, 6g85, 6g86, 6g8i, 6g8j, 6g8k, 6g8l, 6g8p, 6g8q, 6g9b, 6g9i, 6gfx, 6ggb, 6gjj, 6gw1, 6gwe, 6gxe, 6gzl, 6h0b, 6h41, 6h4o, 6h4p, 6h4r, 6h4s, 6h4u, 6h4v, 6h4w, 6h4x, 6h4y, 6h50, 6h51, 6h52, 6h5w, 6h7b, 6h8c, 6h96, 6h9v, 6hck, 6hcu, 6hhp, 6hks, 6hlb, 6hld, 6hle, 6hm4, 6hmg, 6hmt, 6hoi, 6hol, 6hpg, 6hro, 6hs4, 6hv2, 6hy7, 6hza, 6hzb, 6hzc, 6hzd, 6hzx, 6i41, 6i4x, 6i5j, 6i5n, 6i5p, 6i68, 6i7a, 6iae, 6iam, 6idg, 6iiw, 6im4, 6inz, 6iqg, 6iso, 6j9w, 6j9y, 6jad, 6jag, 6jam, 6jan, 6jao, 6jap, 6jax, 6jb0, 6jb4, 6jbb, 6jjm, 6jjn, 6jjz, 6k2n, 6k5r, 6k5t, 6kdi, 6kmj, 6md6, 6me1, 6mil, 6mim, 6min, 6miq, 6mle, 6mlo, 6mm5, 6mnf, 6mqc, 6mqe, 6mqm, 6msy, 6mtv, 6mu3, 6mub, 6n19, 6n3e, 6n3f, 6n5x, 6n7q, 6n87, 6n93, 6nao, 6ncp, 6njz, 6nk0, 6nk1, 6nkp, 6nsx, 6nxz, 6ny0, 6o21, 6o3w, 6o3x, 6o3y, 6o7g, 6oie, 6om2, 6om4, 6oxl, 6p3w, 6p7p, 6p7q, 6pek, 6peu, 6phx, 6pi7, 6pit, 6prg, 6pxc, 6q38, 6q4q, 6q9t, 6q9w, 6qc0, 6qcg, 6qdx, 6qk8, 6qpl, 6qs1, 6qsz, 6qtm, 6qto, 6qtq, 6qtr, 6qts, 6qtw, 6qtx, 6qzr, 6r0x, 6r8i, 6rhe, 6rk4, 6rml, 6rr0, 6s07, 6sen, 6sq0, 6tyz, 6u5m, 6uyx, 6uyy, 6uyz, 6v1c, 7abp, 7kme, 8abp, 9abp, 9icd}


\bibliography{references}
\bibliographystyle{naturemag}